\documentclass[aps,twocolumn,preprintnumbers,prd,superscriptaddress,10pt,nofootinbib]{revtex4-1}

\usepackage{amssymb,amsmath,amsthm,amsfonts}
\usepackage[usenames]{color}
\definecolor{darkred}{rgb}{0.5,0,0}

\usepackage{bm}
\usepackage{dcolumn}
\usepackage[utf8]{inputenc}
\usepackage{latexsym}
\usepackage{rotating}
\usepackage{longtable}

\usepackage{enumerate}
\usepackage{url}
\usepackage[linktocpage]{hyperref}
\usepackage{float}
\usepackage{graphicx}

\def\be{\begin{equation}}
\def\ee{\end{equation}}
\newcommand{\beq}{\begin{eqnarray}}
\newcommand{\eeq}{\end{eqnarray}}

\def\ba{\begin{align}}
\def\ea{\end{align}}

\begin{document}

\title{Strong cosmic censorship: The nonlinear story}

\author{Raimon~Luna}
\affiliation{Departament de F\'{\i}sica Qu\`antica i Astrof\'{\i}sica, Institut
  de Ci\`encies del Cosmos, Universitat de Barcelona, Mart\'{\i} i Franqu\`es 1,
  E-08028 Barcelona, Spain}

\author{Miguel~Zilh\~ao}
\affiliation{CENTRA, Departamento de F\'{\i}sica, Instituto Superior T\'ecnico, Universidade de Lisboa, Avenida Rovisco Pais 1, 1049-001 Lisboa, Portugal}

\author{Vitor~Cardoso}
\affiliation{CENTRA, Departamento de F\'{\i}sica, Instituto Superior T\'ecnico, Universidade de Lisboa, Avenida Rovisco Pais 1, 1049-001 Lisboa, Portugal}
\affiliation{Theoretical Physics Department, CERN, CH-1211 Geneva 23, Switzerland}

\author{Jo\~ao~L.~Costa}
\affiliation{Departamento de Matem\'atica, ISCTE - Instituto Universit\'ario de
  Lisboa, Avenida das Forças Armadas, 1649-026 Lisboa, Portugal}
\affiliation{Center for Mathematical Analysis, Geometry and Dynamical Systems,
  Instituto Superior T\'ecnico, Universidade de Lisboa, Avenida
  Rovisco Pais 1, 1049-001 Lisboa, Portugal}

\author{Jos\'e~Nat\'ario}
\affiliation{Center for Mathematical Analysis, Geometry and Dynamical Systems,
  Instituto Superior T\'ecnico, Universidade de Lisboa, Avenida
  Rovisco Pais 1, 1049-001 Lisboa, Portugal}

\begin{abstract}
A satisfactory formulation of the laws of physics entails that the future evolution of a physical system should be determined from appropriate initial conditions. The existence of Cauchy horizons in solutions of the Einstein field equations is therefore problematic, and expected to be an unstable artifact of General Relativity. This is asserted by the Strong Cosmic Censorship Conjecture, which was recently put into question by an analysis of the linearized equations in the exterior of charged black holes in an expanding universe. Here, we numerically evolve the nonlinear Einstein-Maxwell-scalar field equations with a positive cosmological constant, under spherical symmetry, and provide strong evidence that mass inflation indeed does not occur in the near extremal regime. This shows that nonlinear effects might not suffice to save the Strong Cosmic Censorship Conjecture. 
\end{abstract}

\maketitle

\section{Introduction}
The Strong Cosmic Censorship (SCC) conjecture embodies the expectation that
General Relativity is a deterministic theory. It does so by predicting that  
Cauchy horizons (CH) -- the boundaries of the maximal evolution of initial data via the Einstein field equations -- are unstable, and give rise, upon perturbation, to singular boundaries, beyond which the field equations cease to make sense. 

Quite surprisingly, recent results~\cite{Costa:2017tjc,Costa:2016afl,Hintz:2015jkj,Cardoso:2017soq, Cardoso:2018nvb, Dias:2018etb, Mo:2018nnu,Dias:2018ufh} put into question the validity of SCC in the context of highly charged black holes (BHs) immersed in a spacetime with a positive cosmological constant\footnote{These results appeared almost two decades after the pioneering work in~\cite{Mellor:1990,Brady:1998au}, which made contradicting claims; see section 2.2 in~\cite{Dias:2018etb} for a clarification of the implications of these papers regarding SCC.}. More precisely, these results provide evidence for the existence of stable (charged de Sitter) BH configurations containing a CH in their interior, along which the Misner-Sharp mass, a scalar invariant measuring the energy content of symmetry spheres, remains bounded -- a {\em no mass inflation} scenario. In particular, the CH should then retain enough regularity to allow evolving the spacetime metric across it using the Einstein equations. However, such evolution is not unique, condemning determinism (and the SCC conjecture) to failure.
   
This is particularly disturbing in view of the fact that a positive cosmological constant provides the standard mechanism to model the observed accelerated expansion of our universe. 
Nonetheless, the results mentioned above are restricted to either the linear setting or to the nonlinear analysis of the geometry of the BH interior starting from an already completely formed event horizon -- i.e. the corresponding (horizon) data are put in ``by hand". 
These results provide clear expectations concerning the stability/instability of Cauchy horizons within de Sitter BHs, but are not enough to cast a final verdict on SCC for the following reasons: first, the parameter ranges identified in~\cite{Cardoso:2017soq} as potentially problematic for SCC are very narrow, and therefore even small non-linear deviation form these might be enough to save SCC; secondly the (non-linear) results in~\cite{Costa:2017tjc} do not take into account the oscillatory behavior of the scalar field  along the event horizon\footnote{In~\cite{Costa:2017tjc} it is only considered the case where the (real) scalar field satisfies $\phi_{,v} \sim e^{-kv}$, along the event horizon.} which was identified in~\cite{Cardoso:2017soq} for the first time. Moreover, our numerical simulations will allow us to gain further information concerning the behavior of various relevant quantities near the Cauchy horizon, as for instance a quantitative understanding of tidal deformations.

To go beyond the previous studies we perform a full nonlinear numerical evolution of both massless and massive minimally coupled self-gravitating scalar fields, in a spacetime with a Maxwell field and a positive cosmological constant. For technical reasons we will restrict ourselves to the spherically symmetric setting, but we should stress that, according to the results in~\cite{Cardoso:2017soq}, the spherically symmetric mode plays a key role in the linear stability/instability of Cauchy horizons for near extremal Reissner-Nordstr\"om de Sitter BHs.  

All the nonlinear simulations we will be discussing evolve from characteristic initial data whose outgoing component is located in the BH exterior; in particular, the data along the event horizon arises dynamically in this framework. Our numerical code also allows us to probe the BH interior region and to derive a detailed description of the behavior of fundamental quantities, such as the radius function, scalar field, (Misner-Sharp) mass and curvature. 
By examining the vicinity of the BH parameters identified as potentially problematic for SCC in~\cite{Cardoso:2017soq}, we find stable {\em no mass inflation} scenarios arising from a full nonlinear evolution (of exterior data). These are, to the best of our knowledge, the first results of this kind. They show, in particular, that nonlinear effects are apparently not strong enough to save SCC in the context of highly charged de Sitter BHs\footnote{It should be noted that, in view of recent developments~\cite{Dafermos:2003wr,Luk:2013cqa, Dafermos:2017dbw,Luk:2017jxq,Luk:2017ofx,VandeMoortel:2017ztd}, the situation concerning SCC in the context of asymptotically flat ($\Lambda=0$) BHs  is far more clear.}. 

Our setup also allows the inclusion of a scalar field mass, and so we take the opportunity to investigate the possibility, raised in~\cite{Cardoso:2018nvb} and~\cite{Dias:2018etb}, that the curvature might be bounded up to the Cauchy horizon for certain choices of this mass. This scenario is disproved in the cases that we analyze.

\section{Setup}
We consider here an evolving, electrically charged spacetime,
modelled by the Einstein-Maxwell action with a cosmological constant
$\Lambda$, minimally coupled to a massive scalar field $\Phi$ with mass
parameter $\mu$,
\begin{equation}
S=\int d^4 x {\sqrt {-g}} \left(R-2\Lambda- F^2- 2\Phi_{,\alpha} \Phi^{,\alpha}-2\mu^2 \Phi^2\right) \,, \nonumber
\end{equation}
where $F^2=F_{\alpha \beta} F^{\alpha \beta}$ and $F_{\alpha \beta}$ is the Maxwell tensor.
The equations of motion reduce to
\beq
G_{\mu \nu} + \Lambda g_{\mu\nu} &=& 2F_{\mu \alpha} {F_\nu}^\alpha - \frac{1}{2} g_{\mu \nu} F^2 \nonumber\\
&+&2\Phi_{,\mu} \Phi_{,\nu}- g_{\mu \nu} \left(\Phi_{,\alpha} \Phi^{,\alpha}+\mu^2 \Phi^2 \right)  \, , \label{eq:einstein}\\
\Box \Phi &=& \mu^2 \Phi \,, \qquad dF = d\star\! F=0 \, ,
\eeq
where $\star$ is the Hodge dual.

We focus on spherically symmetric spacetimes, written in double null coordinates as
\beq
\label{eq:metric}
ds^2&=&-2e^{2\sigma(u,v)}du dv + r^2(u,v) d\Omega^2\,, \\
F&=&F_{uv}(u,v)du\wedge dv \,, \qquad  \Phi=\Phi(u,v) \,,
\eeq
where $u$ and $v$ are ingoing and outgoing coordinates, respectively.
In this framework, Maxwell's equations decouple and imply that
\be
F_{u v}r^{2}e^{-2\sigma}= {\rm constant}= Q \,,
\ee
with $Q$ a conserved (electric) charge.

\section{Numerical evolutions}
To numerically evolve the field equations we specify initial conditions along
two null segments, $u=u_i$ and $v=v_i$. We fix the residual gauge freedom as follows:
\be
\label{eq:lingauge}
r(u_i,v) = v\,, \qquad r(u,v_i) = r_0 + u r_{u0}\,,
\ee
where $r_{u0}$ is a constant and $r_0 = v_i$. The profile of the scalar field is set as purely ingoing
\be
\Phi(u_i, v) = A e^{-\left(\frac{v - v_c}{w}\right)^2} \,,
\ee
with the outgoing flux being set to zero, $\Phi_{,u}(u, v_i) = 0$. See the Supplemental Material for more information about the integration procedure.

To interpret our results it will be convenient to consider the following alternative outgoing null coordinates: $\overset{\circ}{v}$, an Eddington-Finkelstein type coordinate, defined by integrating
\begin{equation}
  \label{eq:dvcirc}
  \left(1-\frac{2M}{r} + \frac{Q^2}{r^2} - \frac{\Lambda}{3}r^2\right) d\overset{\circ}{v} = r_{,v} \; dv
\end{equation}
along the event horizon (EH), and $t$, the affine parameter of an outgoing null geodesic, obtained by integrating
\begin{equation}
  \label{eq:dt}
  \left(1-\frac{2M}{r} + \frac{Q^2}{r^2} - \frac{\Lambda}{3}r^2\right) dt
  = - r_{,u}r_{,v} \;  dv
\end{equation}
along a constant $u$ line.  In these expressions $M$ stands for the Misner-Sharp
mass function, which we also closely monitor during the integration, given by
\be
\label{eq:mass}
M(u,v)=\frac{r}{2}\left(1+\frac{Q^2}{r^2}-\frac{\Lambda}{3}r^2 + 2 e^{-2\sigma} r_{, u} r_{, v}\right) \,.
\ee
The constant $r_{u0}$ is thus related to the initial BH mass, $M_0\equiv M(u_i,v_i)$.
Recall that the blow-up of this scalar signals the breakdown of the field equations \cite{Dafermos:2012np} (compare with  \cite{Luk:2013cqa, Dafermos:2017dbw}).

To estimate the curvature we compute the Kretschmann scalar $K$ computed from the field equations (see Supplemental Material; a direct evaluation of this scalar in terms of the metric was found to lead to important round-off error-related problems).

According to the results in Refs.~\cite{Costa:2017tjc,Cardoso:2017soq}, concerning the massless case, we expect the curvature to blow up for all non-trivial initial data throughout the entire subextremal parameter range. Although it is a potentially interesting nonlinear effect, we recall that the blow-up of $K$, per se, is of little significance: it implies neither the breakdown of the field equations~\cite{L2}
nor the destruction of macroscopic observers~\cite{Ori:1991zz}.
Recall that the results in Refs.~\cite{Cardoso:2018nvb,Dias:2018etb} suggest that the introduction of scalar mass could lead, for appropriate choices of BH parameters, to solutions with bounded curvature. As we will see below, our results contradict this expectation.

\section{Initial conditions}
The physical problem is then fully determined upon specifying $Q$, $\Lambda$,
$\mu$, $M_0$, $A$, $v_c$ and $w$. Since our purpose here is to determine whether the
linearized predictions of Refs.~\cite{Cardoso:2017soq,Cardoso:2018nvb} hold in the full nonlinear
regime, we focus on $M_0=1,\,\Lambda=0.06$ and use the following configurations:

\noindent {\bf A}: $Q=0.9000,\,\mu = 0$, corresponding to $Q=0.890Q_{\rm max}$. In this case, the results in Ref.~\cite{Cardoso:2017soq} (lower left panel of Fig.~3) predict {\em mass inflation}.

\noindent {\bf B}: $Q=1.0068,\,\mu = 0$, corresponding to $Q=0.996Q_{\rm max}$. Linearized studies provide evidence in favor of a {\em no mass inflation} scenario~\cite{Cardoso:2017soq}.

\noindent {\bf C}: $Q=1.0068,\,\mu = 1.0$. The results of Ref.~\cite{Cardoso:2017soq} together with those of Ref.~\cite{Cardoso:2018nvb} -- see Fig.~2 -- also provide evidence in favor of a {\em no mass inflation} scenario.
Here we are considering the superposition of both neutral massless scalar perturbations~\cite{Cardoso:2017soq} and charged massive scalar perturbations~\cite{Cardoso:2018nvb} as being the most predictive of the full non-linear evolution. If we just take into account massive scalar perturbations, then the results in Ref.~\cite{Cardoso:2018nvb} (see Fig.~2]) and~\cite{Dias:2018etb} (page 22) suggest that curvature might also be bounded.

To test the dependence of our results on initial data, we use the following initial profiles for the scalar
field:

\noindent ${\bf 1}$: $A = 0.04,\,w = 0.1$ and $v_c = 3.0$;

\noindent ${\bf 2}$: $A = 0.08,\,w = 0.5$ and $v_c = 3.0$.

\begin{figure}[thb]
  \includegraphics[width=0.5\textwidth]{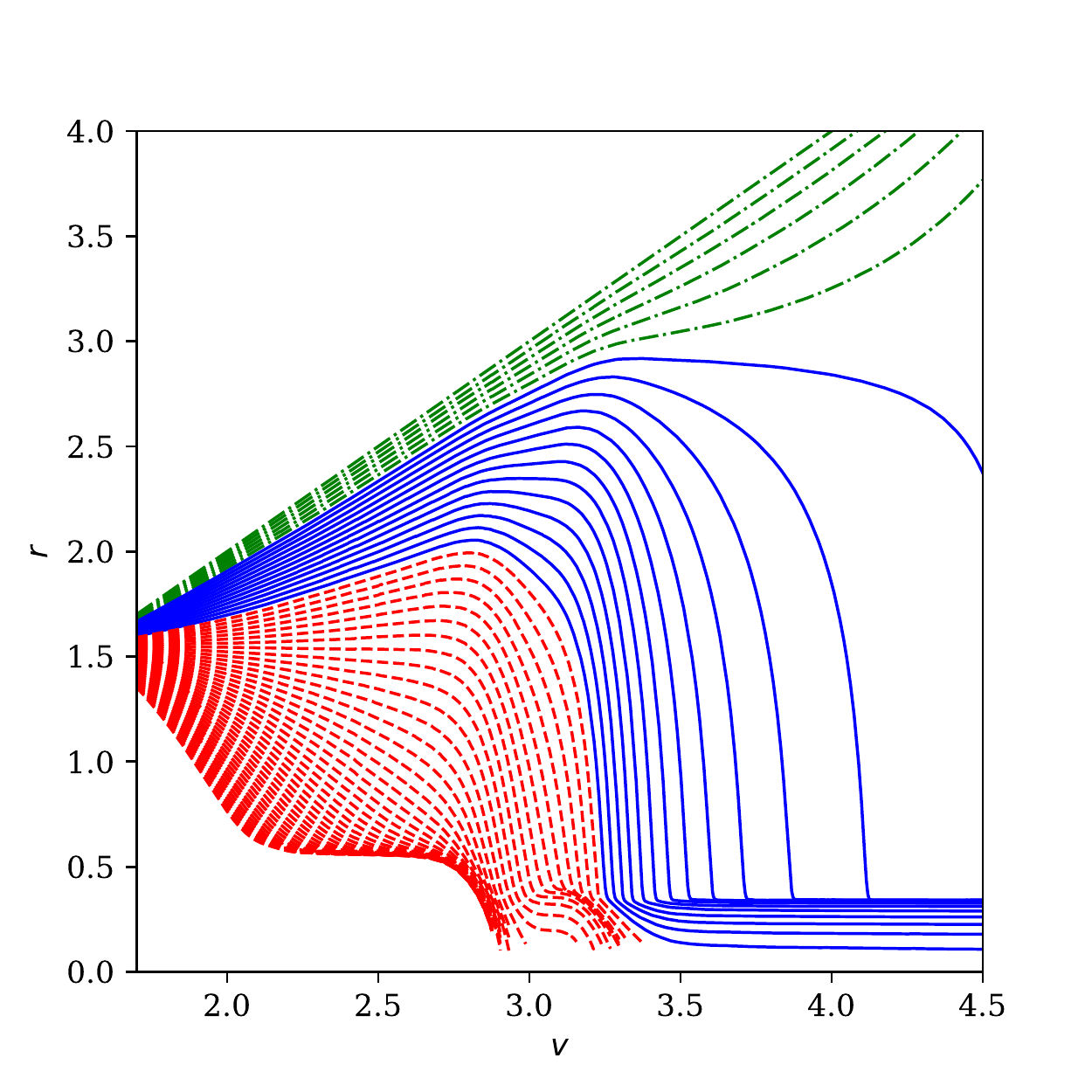}
  \caption{Radius function for constant-$u$ slices in a configuration with
    $M_0 = 1.0$, $Q = 0.9$, $\Lambda = 0.06$, $\mu = 0$, $A = 0.4$, $v_c = 3.0$
    and $w = 0.25$. Dashed-dotted green lines reach infinity, full blue lines
    hit the CH and red dashed lines hit the singularity at $r=0$. \label{fig:radius}}
\end{figure}
We have evolved the relevant system of equations using the \textsc{DoNuTS} code,
described briefly in the Supplemental Material. It is based on the formulation
presented in Refs.~\cite{Burko:1997tb,Hansen:2005am,Avelino:2011ee}, but the 
integration technique makes it spectrally accurate in the
$v$-direction and, correspondingly, runs with trivial memory requirements and
orders of magnitude faster than previously reported codes. 

\section{Results}
%
\begin{figure}[thbp]
\includegraphics[width=0.5\textwidth]{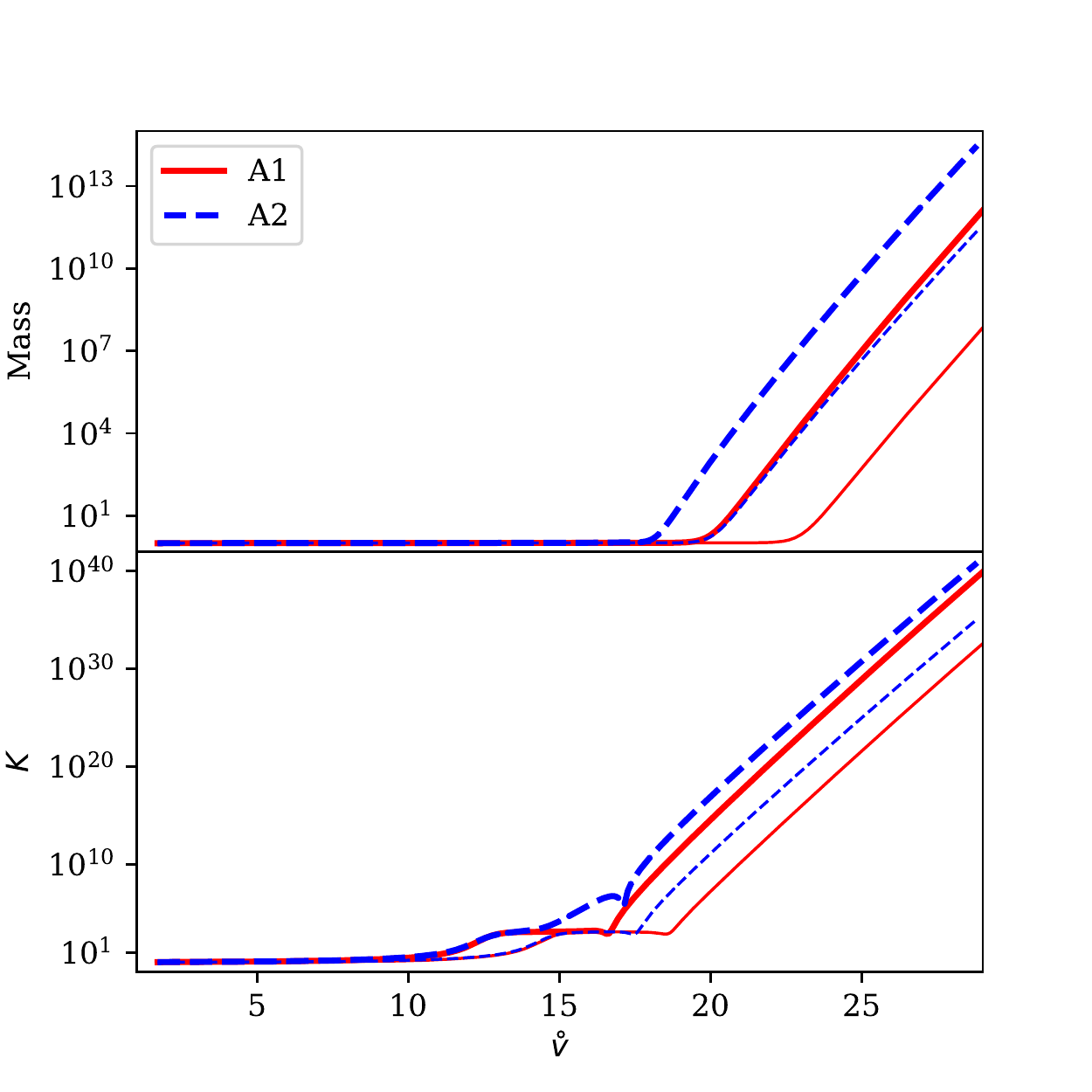}
\caption{Mass function~\eqref{eq:mass} and Kretschmann scalar as functions of $\overset{\circ}{v}$ for configurations {\bf A1} (red solid line) and
{\bf A2} (blue dashed line). Thin lines are evaluated at $u=u_{\rm EH} + 1$ and thick lines are evaluated at $u=u_{\rm EH} + 2$.
These results are consistent with the existence of mass inflation leading to a weak singularity.
\label{fig:inflation}}
\end{figure}
It is important to start by noticing that, as widely expected~\cite{Costa:2017tjc,Dafermos:2003wr}, our numerics show that all solutions contain a non-empty CH in their BH interior. This can be attested by monitoring the radius function -- shown in Fig.~\ref{fig:radius} -- along null lines $u=u_1$, for $u_1>u_{\rm EH}$, where $u=u_{\rm EH}$ is the event horizon. In fact, for  $u_1$ larger but close to $u_{\rm EH}$, the radius converges, in $v$, to a non-vanishing constant. It is also interesting to note that, for some initial configurations and large enough $u_1$, the radius does converge to zero, signaling (in that region) a singularity beyond which the metric cannot be extended \cite{Sbierski:2015nta}.

As is well known the behavior of the scalar field along the event horizon is of great significance for the structure of the BH interior region. The first noteworthy feature of our results (see Supplemental Material for details) is that,  as expected, the field  decays exponentially (in $\mathring{v}$). More surprisingly, we also clearly observe an oscillatory profile; this might seem odd at first, since it is in contrast with what happens for $\Lambda=0$ and with the expectation created by the study of sufficiently sub-extremal BHs with $0<M^2\Lambda\ll 1$~\cite{Brady:1996za}. However, it turns out that such behavior should be expected from the linearized analysis of Refs.~\cite{Cardoso:2017soq,Cardoso:2018nvb}, where it is shown that, for a configuration resembling our configuration {\bf B}, there are two modes which dominate the response: a non-oscillatory ``near extremal'' (NE) mode with characteristic frequency $\omega_{\rm NE}\sim-0.081 i$, {\it and} a ``photon sphere'' (PS) mode with $\omega_{\rm PS}\sim 0.096 - 0.095i$ (these numbers are given in the units and time-coordinate of Ref.~\cite{Cardoso:2017soq}). Here we find very good agreement with the PS mode (when translated to our $\mathring{v}$ coordinate) which is oscillatory in nature. Similar agreement can be found for the remaining configurations {\bf A} and {\bf C}. We also recall that, according to the results in~\cite{Cardoso:2017soq}, in the $M^2\Lambda\ll 1$ case, the dominant mode is a non-oscillatory ``de Sitter'' mode, in agreement with~\cite{Brady:1996za}.  

\begin{figure}[thbp]
\includegraphics[width=0.5\textwidth]{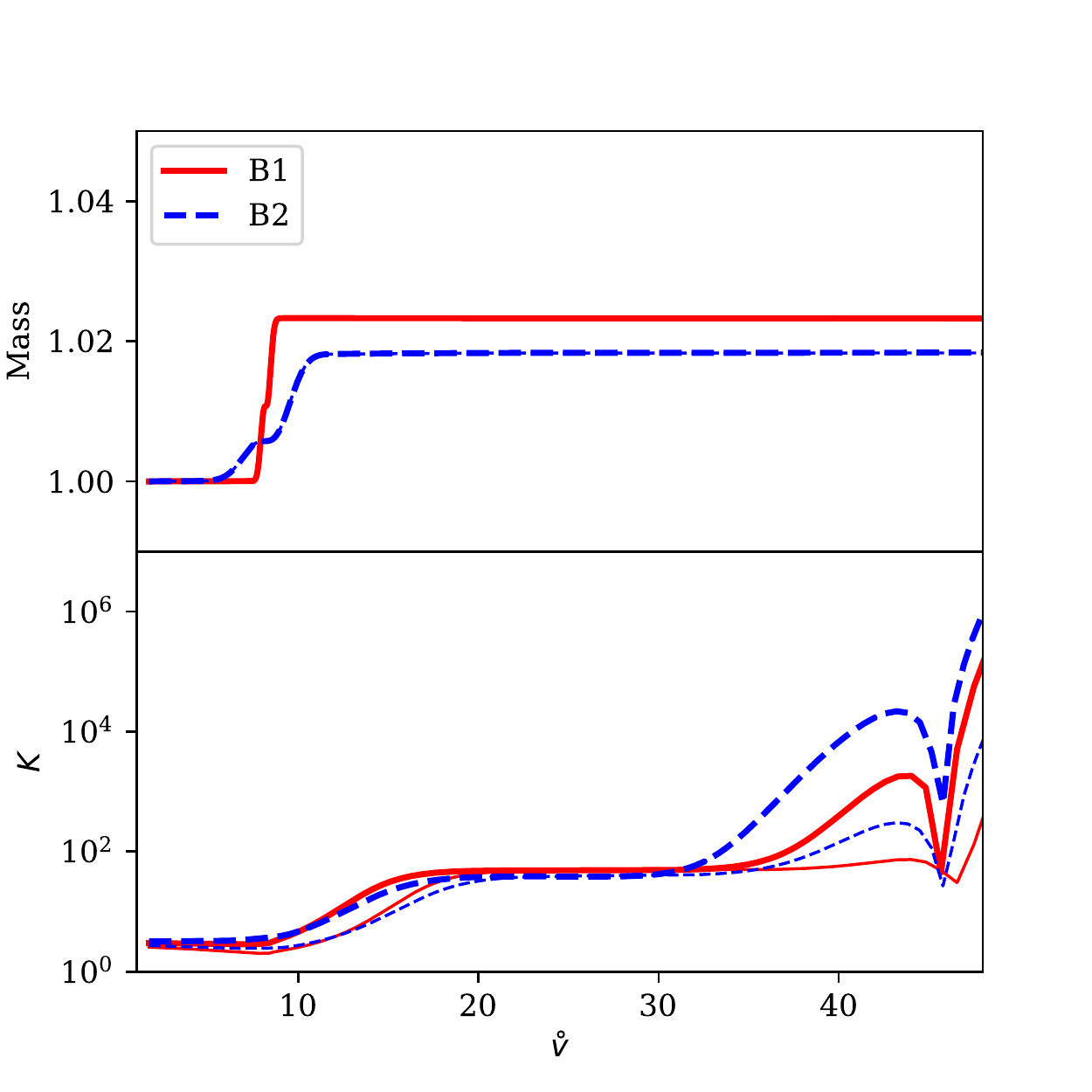}
\caption{Mass function~\eqref{eq:mass} and Kretschmann scalar
as functions of $\overset{\circ}{v}$ for configurations {\bf B1} (red solid line) and
{\bf B2} (blue dashed line). Thin lines are evaluated at $u=u_{\rm EH} + 1$ and
thick lines are evaluated at $u=u_{\rm EH} + 2$.
\label{fig:no_inflation}}
\end{figure}

\begin{figure}[thbp]
\includegraphics[width=0.5\textwidth]{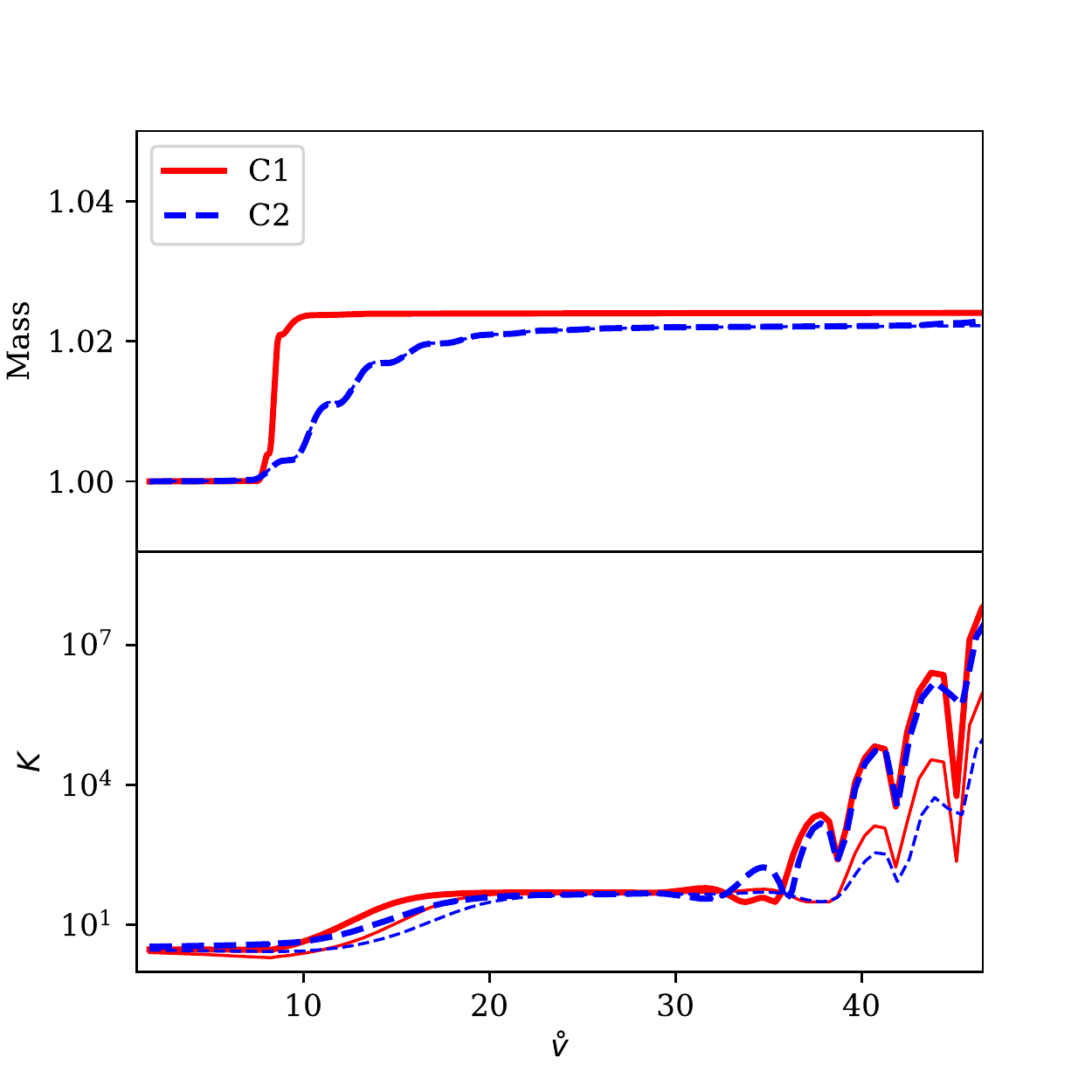}
\caption{Same as Fig.~\ref{fig:no_inflation}, for configurations {\bf C1} (red solid) and
{\bf C2} (blue dashed).
\label{fig:massive}}
\end{figure}

Our main results (concerning mass and curvature) are summarized in Figs.~\ref{fig:inflation}-\ref{fig:massive}.
Fig.~\ref{fig:inflation} shows the evolution of the mass function and the Kretschmann scalar for configurations {\bf A}: in these cases mass inflation occurs, and, consequently, the curvature invariant $K$ diverges. Note that an observer crossing one such region will be subjected to physical {\it deformations} which are not necessarily infinite (see discussion below). 
Nonetheless, because there is mass inflation, the singularity is strong enough to deserve the classification of {\em terminal boundary}, since it corresponds to a locus where the field equations cease to make sense. These conclusions are consistent with the linear results in~\cite{Cardoso:2017soq} and the nonlinear results in~\cite{Costa:2017tjc}.

The main novel result of this work concerns Fig.~\ref{fig:no_inflation}: there exist configurations for which no mass inflation occurs. For configurations {\bf B}, in the BH interior, after a small ``accretion'' stage the mass settles to a constant value. Moreover, as recently predicted~\cite{Costa:2017tjc,Cardoso:2017soq}, the CH remains a  curvature singularity, since the curvature scalar $K$ diverges. 
However, the lack of mass inflation makes the singularity so ``mild'' that, in principle, one should be able to continue the evolution of the space-time metric across it, by solving the Einstein field equations!

A somewhat unexpected feature (of configuration {\bf B}) is the oscillatory way in which the curvature scalar diverges. In hindsight, such behavior could be expected from the previously discussed oscillatory behavior of the scalar field along the event horizon. Note that in a no mass inflation situation it is the blow up of $\Phi_{,v}/r_{,v}$ that dominates the behavior of $K$. This should be contrasted with what happens when mass inflation occurs: then it is the {\em monotone} divergence of the mass that controls the Kretschmann; this last fact provides an explanation for the non-oscillatory behavior observed for configuration {\bf A}.

Concerning massive scalars, the results presented in Fig.~\ref{fig:massive} identify configuration {\bf C} as another no mass inflation configuration. Once again we find that the corresponding CH is a ``weak'' curvature singularity. In fact, the presence of scalar mass seems to have no attenuation effect on the growth of $K$, in contrast with what might be expected from linear analysis~\cite{Cardoso:2018nvb} and~\cite{Dias:2018etb}.

We finish this section with some further remarks concerning the blow up of curvature. In configuration {\bf A}, our results indicate that the Kretschmann scalar blows up as $t^{-2}$ (possibly modulated
by logarithmic terms), where $t$ is the affine parameter defined in \eqref{eq:dt} with the Cauchy horizon located at $t=0$. This might suggest that the curvature blows up as $t^{-1}$, but, as noted in~\cite{Poisson:1989zz,Ori:1991zz}, there are curvature components that may blow up even faster. In fact, the quantities that determine the blow-up of the Kretschmann scalar are the square of the (Misner-Sharp) mass $M$ and the square of the gradient of $\Phi$, which is dominated by $(\Phi_{,v}/r_{,v})^2$. However, all curvature components are controlled by $M$ and $(\Phi_{,v}/r_{,v})^2$ (the origin of the last term can be traced to the energy-momentum tensor). From the behavior of the Kretschmann scalar we can then conclude that the components of the curvature blow up at most as $t^{-2}$; we also expect inverse logarithmic powers~\cite{Ori:1991zz, Luk:2013cqa, Dafermos:2017dbw} that are hard to detect numerically. Although divergent, these curvature components should yield (with the help of the logarithmic terms) a finite ``tidal deformation'' when integrated twice with respect to $t$, in agreement with the picture in~\cite{Ori:1991zz}.


From the equation (see Ref.~\cite{Costa:2014aia})
\be
M_{,v} = \frac12 \left(1-\frac{2M}{r} + \frac{Q^2}{r^2} - \frac{\Lambda}{3}r^2\right) \left(\frac{\Phi_{,v}}{r_{,v}}\right)^2 r_{,v}
\ee
we conclude that no mass inflation is essentially equivalent to the integrability of $(\Phi_{,v}/r_{,v})^2$, with respect to $t$\footnote{In fact, there is a mathematical equivalence under the reasonable, in principle generic, assumption that the quantity $1-\frac{2M}{r} + \frac{Q^2}{r^2} - \frac{\Lambda}{3}r^2$ does not vanish at the Cauchy horizon.}. Moreover, when both occur we immediately see that the curvature can only give rise to finite ``tidal deformations''.   This reasoning is verified by our results concerning configuration {\bf B}, for which the mass is bounded,  $\Phi_{,v}/r_{,v}$ grows slower than $t^{-1/2}$ and the Kretschmann scalar and the curvature components blow up at most as $t^{-1}$.

\section{Discussion}
The main motivation for this study was to understand whether nonlinear effects could trigger mass inflation, even when the linearized analysis suggests otherwise~\cite{Cardoso:2017soq}. We found that nonlinear effects are not strong enough to change the picture: in fact, the nonlinear results are in full agreement with the linearized predictions. The linearized analysis of Ref.~\cite{Cardoso:2017soq} suggests that no mass inflation should occur for BH charge above a threshold $Q_{*}\simeq 0.95$. Within a nonlinear evolution, the precise linearized results are difficult to reproduce (for instance, the final spacetime parameters depend on the initial parameters and on the size of the initial data).
However, for small scalar amplitudes, our results are indeed consistent with the previous threshold.

The (numerical) no mass inflation solutions presented here are the first solutions of this kind arising from the full nonlinear evolution of exterior data. They contain a Cauchy horizon in their BH interior region that can be seen as (``weakly'') singular, due to the divergence of curvature invariants. However, these divergent tidal forces are not necessarily strong enough to lead to a divergent tidal deformation and the consequent unequivocal destruction of all macroscopic objects~\cite{Ori:1991zz}. Even more problematic, the lack of mass inflation indicates that these Cauchy horizons should maintain enough regularity as to allow the field equations to determine (classically), in a highly non-unique way, the evolution of the metric to their future. This corresponds to a potential severe violation of SCC.

Our results concern spherically symmetric spacetimes. The picture is unlikely to change
even with asymmetric initial conditions~\cite{Cardoso:2017soq}. Thus, from the {\it conceptual} point of view~\cite{Cardoso:2018nvb},
our results show that SCC is not enforced by the field equations. 
In the meantime, interesting suggestions to remedy SCC, in the presence
of a positive cosmological constant, have been put forward: these include enlarging the allowed
set of initial data by weakening their regularity~\cite{Dafermos:2018tha}, or restricting the scope of SCC to the uncharged BH  setting~\cite{Dias:2018ynt}.
It thus seems plausible that the {\it astrophysical} interpretation of SCC remains valid, once other fields {\it and} realistic BH charges are considered.

\begin{acknowledgments}

We are indebted to Kyriakos Destounis, Aron Jansen, and Peter Hintz for many and very useful conversations and comments.
R.~L. gratefully acknowledges the hospitality of the group at CENTRA, where this work was started.
We are grateful to the Yukawa Institute for Theoretical
Physics at Kyoto University, for hospitality while this work
was completed during the YITP-T-18-05 on ``Dynamics in
Strong Gravity Universe.''
R.~L. is supported by Ministerio
de Educación, Cultura y Deporte Grant No. FPU15/01414.
M.~Z. acknowledges financial support provided by FCT/Portugal through the IF program, Grant No. IF/00729/2015. 
J.~L.~C. and J.~N. acknowledge financial support
provided by FCT/Portugal through UID/MAT/04459/2013 and Grant No. (GPSEinstein) PTDC/MAT-ANA/1275/2014. 
The authors acknowledge financial support
provided under the European Union's H2020 ERC
Consolidator Grant ``Matter and strong-field gravity: New frontiers in Einstein’s theory'' Grant No. MaGRaTh-646597. 
This project has received funding from the
European Union's Horizon 2020 research and innovation
program under the Marie Sklodowska-Curie Grant
No. 690904. 
The authors would like to acknowledge
networking support by the GWverse COST Action
CA16104, ``Black holes, gravitational waves and fundamental physics.'' 
This research was supported in part by
Perimeter Institute for Theoretical Physics. Research at
Perimeter Institute is supported by the Government of
Canada through the Department of Innovation, Science,
and Economic Development, and by the Province of
Ontario through the Ministry of Research and Innovation.
\end{acknowledgments}

\appendix

\section{Numerical procedure}

\subsection{Algorithm}

Our equations of motion have the form
\begin{align}
  r_{,uv} + \frac{r_{,u}r_{,v}}{r}
  + \frac{e^{2\sigma}}{2r}\left[1- \frac{Q^2}{r^2} - \left(\Lambda + \mu^2 \Phi^2\right) r^2\right]
  = 0\,, \label{eq:revol}\\
  \sigma_{,uv} - \frac{r_{,u}r_{,v}}{r^2}
  - \frac{e^{2\sigma}}{2r^2}\left(1- 2\frac{Q^2}{r^2}\right)
  + \Phi_{,u}\Phi_{,v}= 0 \,, \label{eq:sigmaevol}\\
  \Phi_{,uv} + \frac{1}{r}\left(\Phi_{,u}r_{,v} + \Phi_{,v}r_{,u}\right)  + \frac{e^{2\sigma}}{2}\mu^2\Phi = 0 \,, \label{eq:phievol}
\end{align}
and are subjected to the following constraints
\begin{align}
 r_{,uu} - 2r_{,u}\sigma_{,u} + r\,(\Phi_{,u})^2 = 0 \,, \label{eq:constraintu}\\
 r_{,vv} - 2r_{,v}\sigma_{,v} + r\,(\Phi_{,v})^2 = 0 \,. \label{eq:constraintv}
\end{align}
These equations must be satisfied by the initial data. Then, by virtue of the
Bianchi identities, they will be satisfied in the whole computational domain
provided that the dynamical equations are accurately satisfied.

To integrate these equations, we start by transforming them into a system of
ODEs. Our procedure is as follows. Let $h(u,v)$ be any evolved quantity
$r(u,v)$, $\sigma(u,v)$ and $\Phi(u,v)$.
Defining $f(v) = \partial_{u} h(u, v)$, all dynamical equations, \emph{for fixed
  $u$}, have the form
\begin{equation}
\label{eq:fp}
f'(v) + f(v)p(v) = g(v)\,,
\end{equation}
where $'$ denotes the derivative with respect to $v$. These equations can be
solved by introducing the integrating factor
\[
  \lambda(v) = \exp\left(\int_{v_i}^v p(v') \, dv'\right), \qquad
  \lambda'(v) = p(v)\lambda(v) \,.
\]
Multiplying Eq.~(\ref{eq:fp}) by $\lambda(v)$, we get
\begin{align*}
  f'(v)\lambda(v) + f(v)\lambda'(v)  = \left[f(v)\lambda(v)\right]'
  = g(v)\lambda(v) \\
  \Leftrightarrow f(v) \equiv \partial_u h(v) = \frac{1}{\lambda(v)}
  \left[f(v_i) + \int_{v_i}^v g(v')\lambda(v') \, dv'\right]\,,
\end{align*}
which are ODEs in $u$ for all values of $v$.  Given initial conditions in the
two null segments $u = u_i$, $h(u_i, v)$ $\forall v$ and $v = v_i$,
$f(v_i) \equiv \partial_u h(u,v_i)$ $\forall u$, we can integrate the equations in a
rectangular region $u_i < u < u_f$ and $v_i < v < v_f$.

For our three functions in Eqs.~(\ref{eq:revol}), (\ref{eq:sigmaevol}) and
(\ref{eq:phievol}), $p(v)$ and $g(v)$ are the following:
\begin{align*}
  p_r(v) & = \frac{r_{,v}}{r}\,, \\
  g_r(v) & = -\frac{e^{2\sigma}}{2r}\left[1- \frac{Q^2}{r^2}
           - \left(\Lambda + \mu^2 \Phi^2\right) r^2\right]\,, \\
  p_{\Phi}(v) & = \frac{r_{,v}}{r}\,, \\
  g_{\Phi}(v) & = -\frac{r_{,u} \Phi_{,v}}{r}
                - \frac{e^{2\sigma}}{2}\mu^2\Phi\,, \\
  p_{\sigma}(v) & = 0 \,, \\
  g_{\sigma}(v) & = \frac{r_{,u} r_{,v}}{r^2}
                  +\frac{e^{2\sigma}}{2r^2}\left(1-2\frac{Q^2}{r^2}\right)
                  - \Phi_{,u} \Phi_{,v} \,.
\end{align*}
We integrate these equations using the \emph{Double Null Through Spectral
  methods} (\textsc{DoNuTS}) code written in Julia~\cite{bezanson2017julia}. To
integrate the system within \textsc{DoNuTS}, all functions are expanded in a
Chebyshev basis in the $v$ direction (where all $v$ derivatives and integrations
can be readily performed), and the remaining ODEs in the $u$ direction are
integrated using an adaptive step integrator through the
DifferentialEquations.jl Julia
package~\cite{rackauckas2017differentialequations}.

\subsection{Adaptive gauge}

When using the initial gauge, $r_{, u}$
becomes extremely large around the apparent horizon for large $v$. Therefore, in
order to explore the near-horizon region at late times, 
it is convenient to use an adaptive gauge in $u$
during the numerical evolution.

Since the change $u \to \tilde{u}(u)$ together with $\sigma \to \sigma - \frac{1}{2} \log\left(\frac{d\tilde{u}}{du}\right) $ leaves the equations invariant, we can change the gauge in $u$ along the integration by choosing appropriately the initial condition $\sigma_{,u}(u,v_i)$ at each value of $u$.

To explore the near-horizon geometry, we can choose an Eddington-like gauge for $u$, i.e., a gauge that brings the event horizon to $u \to \infty$. A good way to do so, as described in~\cite{Eilon:2015axa}, is to set $\sigma(u, v_f) = \log\left(2r_{,v}(u, v_f)\right) + C$, where $C$ can be any constant. In the \textsc{DoNuTS} code, this is achieved by picking the initial condition for $\sigma_{,u}(u,v_i)$ 
\begin{equation}
\label{eq:adaptgaugesigma}
\sigma_{,u}(u, v_i) = - \left[\sigma(u, v_f) - \log\left(2r_{,v}(u, v_f)\right) + \frac{3}{2}\log 2 \right] \,.
\end{equation}
The term $\frac{3}{2}\log 2$ is chosen so that $\sigma_{,u}(u_i, v_i)$ is small when $\sigma(u_i, v_f) \approx -\frac{1}{2}\log 2$. With this condition, $\sigma(u, v_f)$ is damped towards the desired value $\log\left(2r_{,v}(u, v_f)\right) - \frac{3}{2}\log 2$ along the evolution in $u$. Additionally, in order to satisfy the constraint equation (\ref{eq:constraintu}), we must introduce an additional ODE for the initial condition $r_{,u}(u, v_i)$ at $v = v_i$,
\begin{equation}
\label{eq:adaptgauger}
r_{,uu}(u, v_i) = 2r_{,u}(u, v_i)\sigma_{,u}(u, v_i)
\end{equation}
with $r_{,u}(u_i, v_i) = r_{u0}$ obtained using the expression for the Misner-Sharp mass in the main text. By solving this ODE along with all the others, we get the initial conditions at $v = v_i$ at each value of $u$ along the integration.

\begin{figure*}[tbph]
\includegraphics[width=0.45\textwidth]{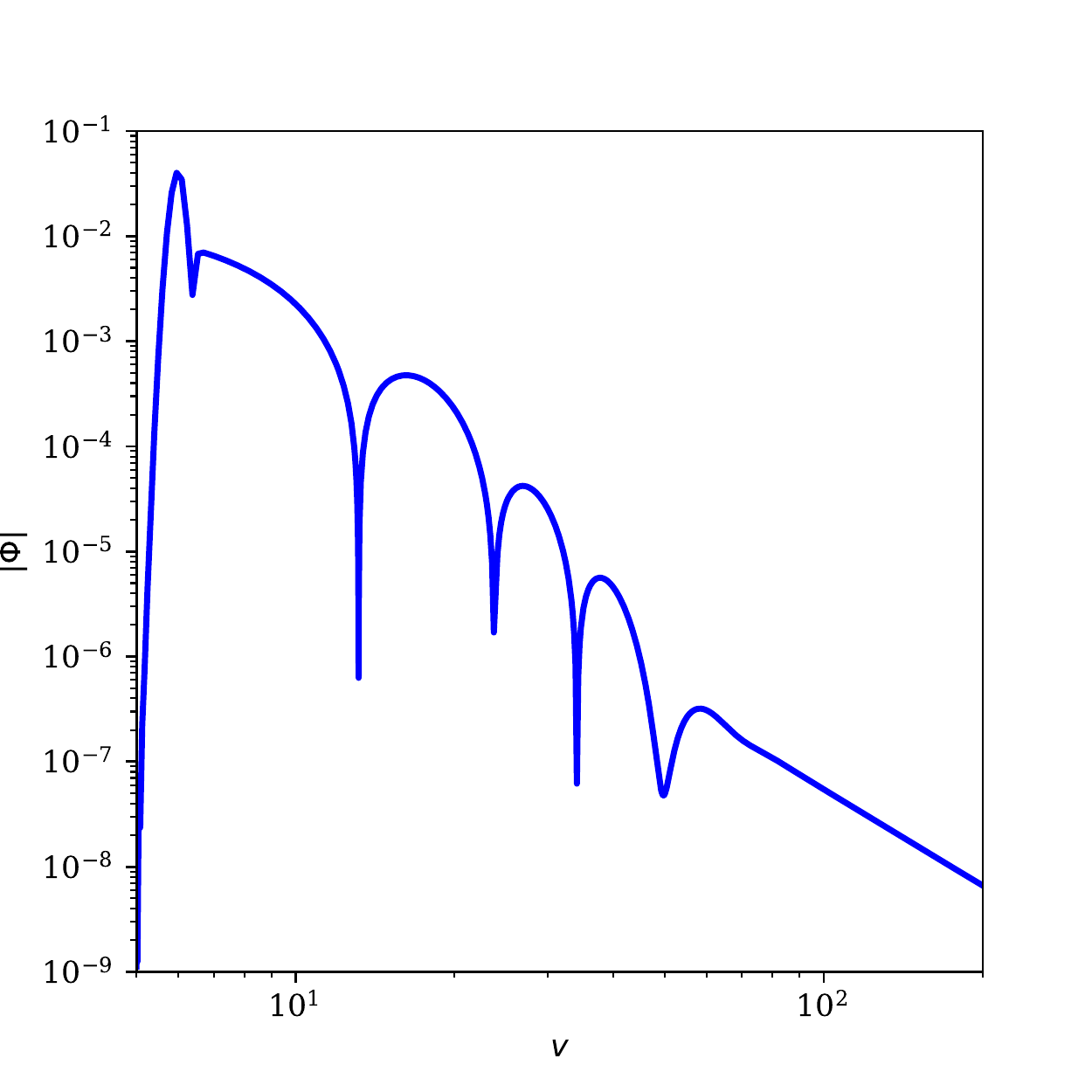} \hfill
\includegraphics[width=0.45\textwidth]{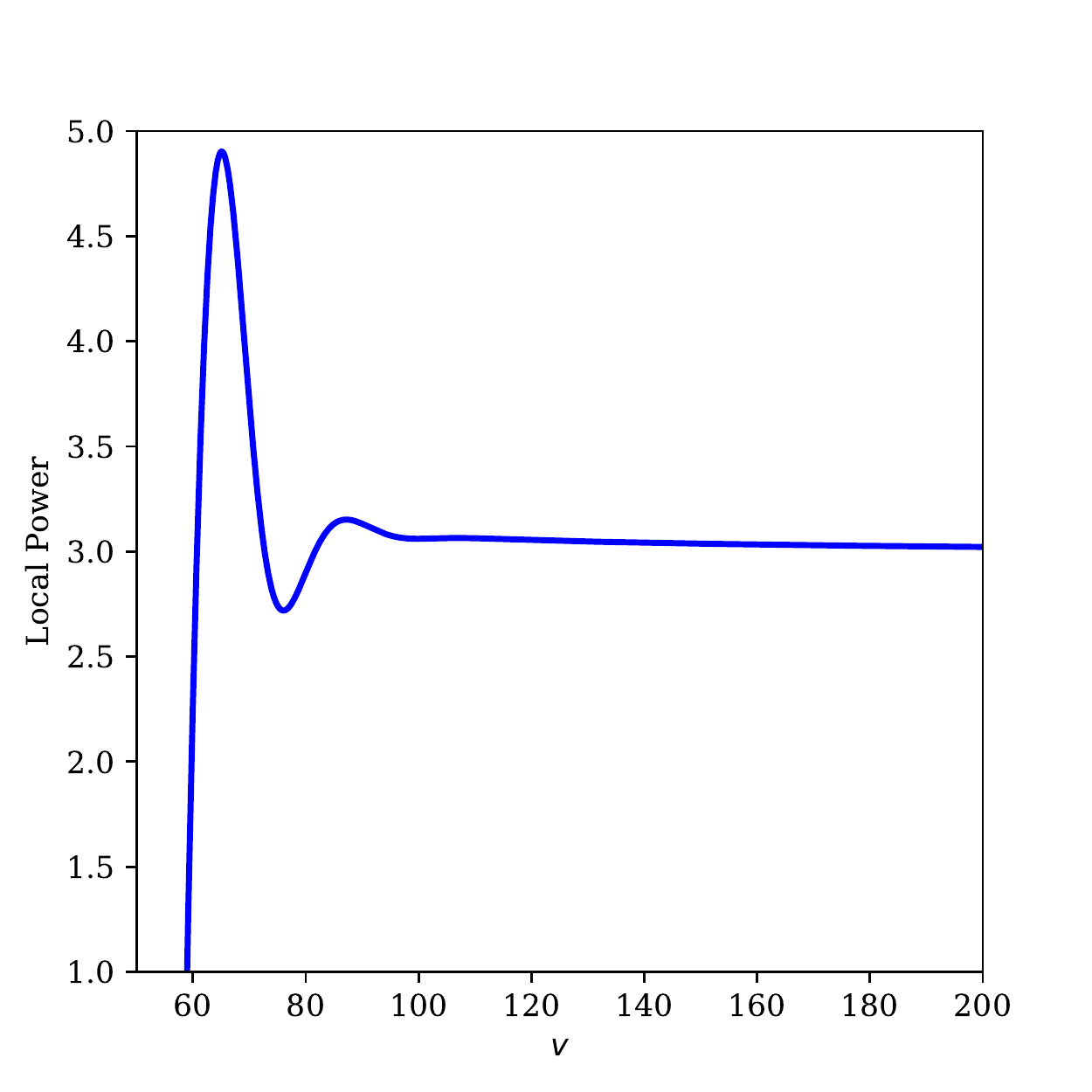}
\caption{Massless scalar field along the event horizon with corresponding ``local power'' for a configuration with $M_0 = 1.0$, $Q = 0.95$, $\Lambda = 0$, $\mu = 0$, $A = 0.01$, $v_c = 6.0$ and
$w = 0.25$. The power-law decay $\Phi \sim v^{-3}$ matches to a very good precision the one expected from linearized analysis~\cite{Price:1971fb}, and reproduces well previous nonlinear results~\cite{Burko:1997tb}.\label{fig:latetime}}
\end{figure*}
%
\subsection{Code tests}
As a test of our numerical implementation we have reproduced the late-time decay
of an asymptotically flat configuration with a massless scalar field. For this, it was crucial to employ the gauge
conditions~(\ref{eq:adaptgaugesigma}), (\ref{eq:adaptgauger}). We also compute
the ``local power'' of the scalar field decay, defined as
$-v\Phi_{,v}/\Phi$. These are shown in Fig.~\ref{fig:latetime} and are
consistent with expected results.

To further test the code, we have analyzed its convergence properties.  We thus
evaluate the quantity
\begin{equation}
\label{eq:err}
\delta_{n,m} (F) \equiv \max|1-F_n/F_m|
\end{equation}
for a given function $F_N$ obtained with resolution $N$ at a fixed $u$
coordinate, and where the maximum is evaluated for all values of $v$.
Here, the index $m$ refers to a reference solution obtained using a large
number $m$ of grid points while $n$ denotes test solutions using a coarser
resolution, $n < m$.

\begin{figure}[thbp]
  \includegraphics[width=0.47\textwidth]{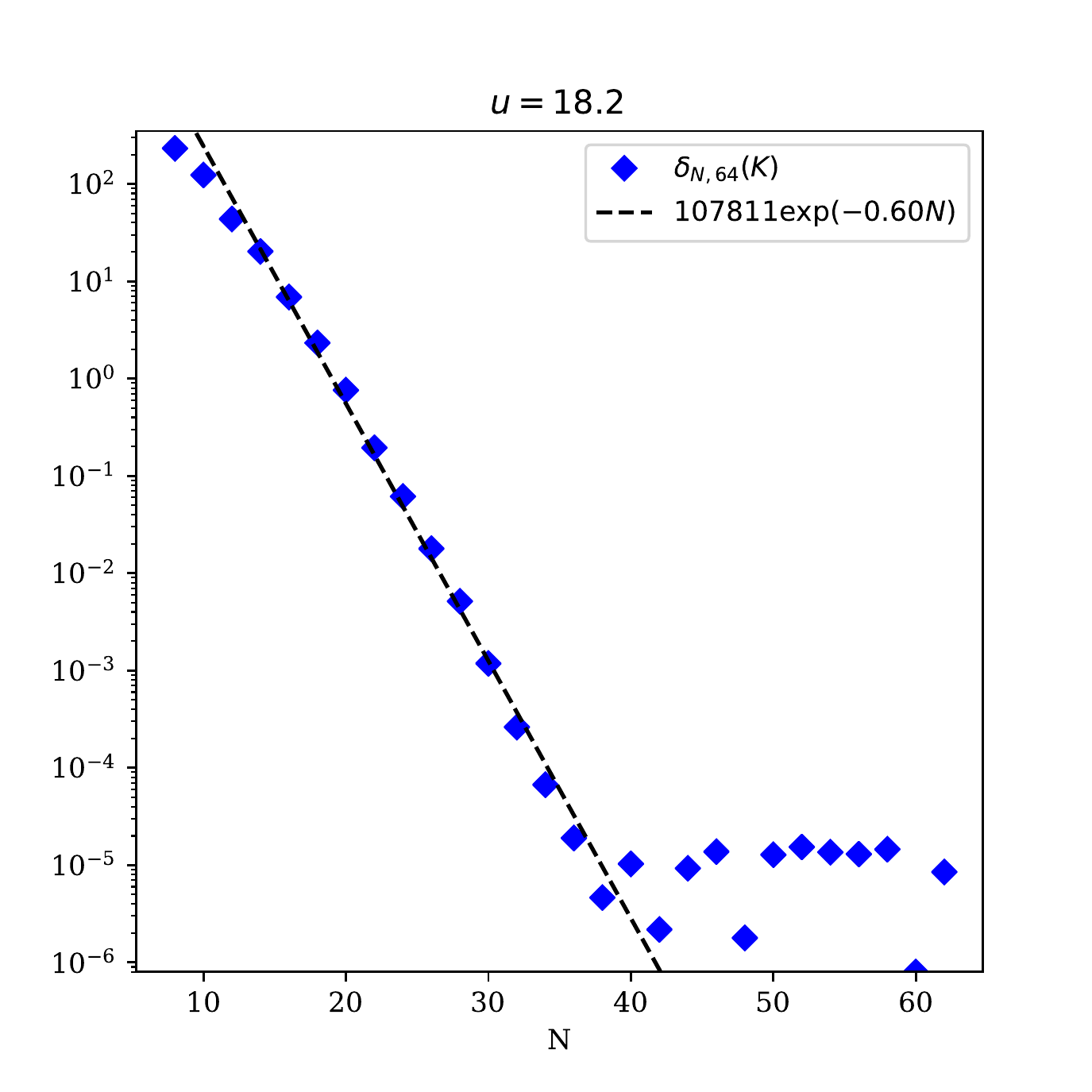}
  \caption{$\delta_{N,64} (K)$ at $u=18.2$ for configuration {\bf B1}.
    20 domains were employed in the $v$ direction, where each domain has $N$
    points. The plot clearly shows exponential convergence until
    $N\approx40$. \label{fig:conv} }
\end{figure}

In Fig.~\ref{fig:conv} we show the convergence properties of the Kretschmann
scalar for configurations {\bf B1}. The plots show exponential convergence up to
$N \approx 40$.

\begin{figure*}[thbp]
  \includegraphics[width=0.48\textwidth]{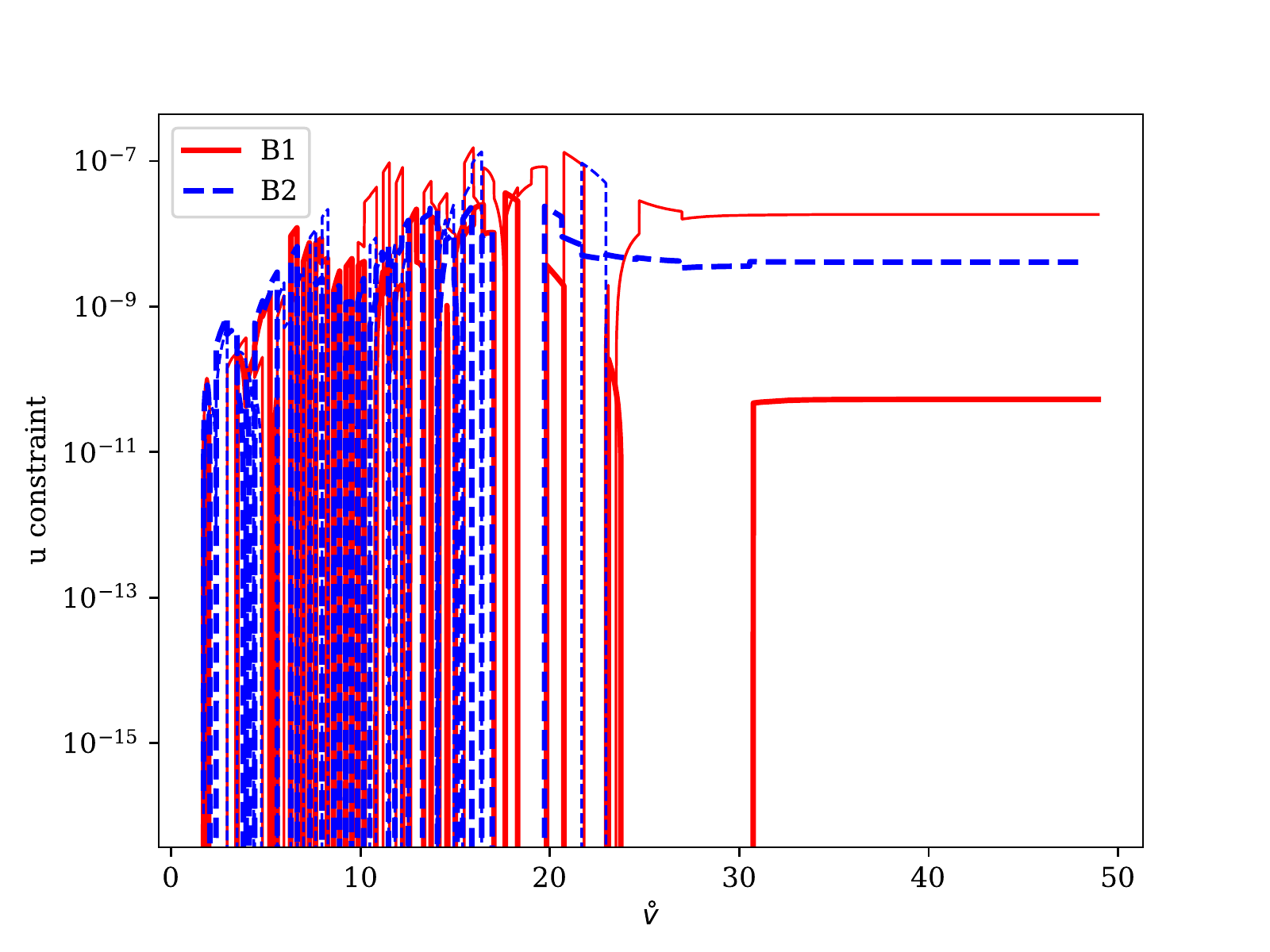} \hfill
  \includegraphics[width=0.48\textwidth]{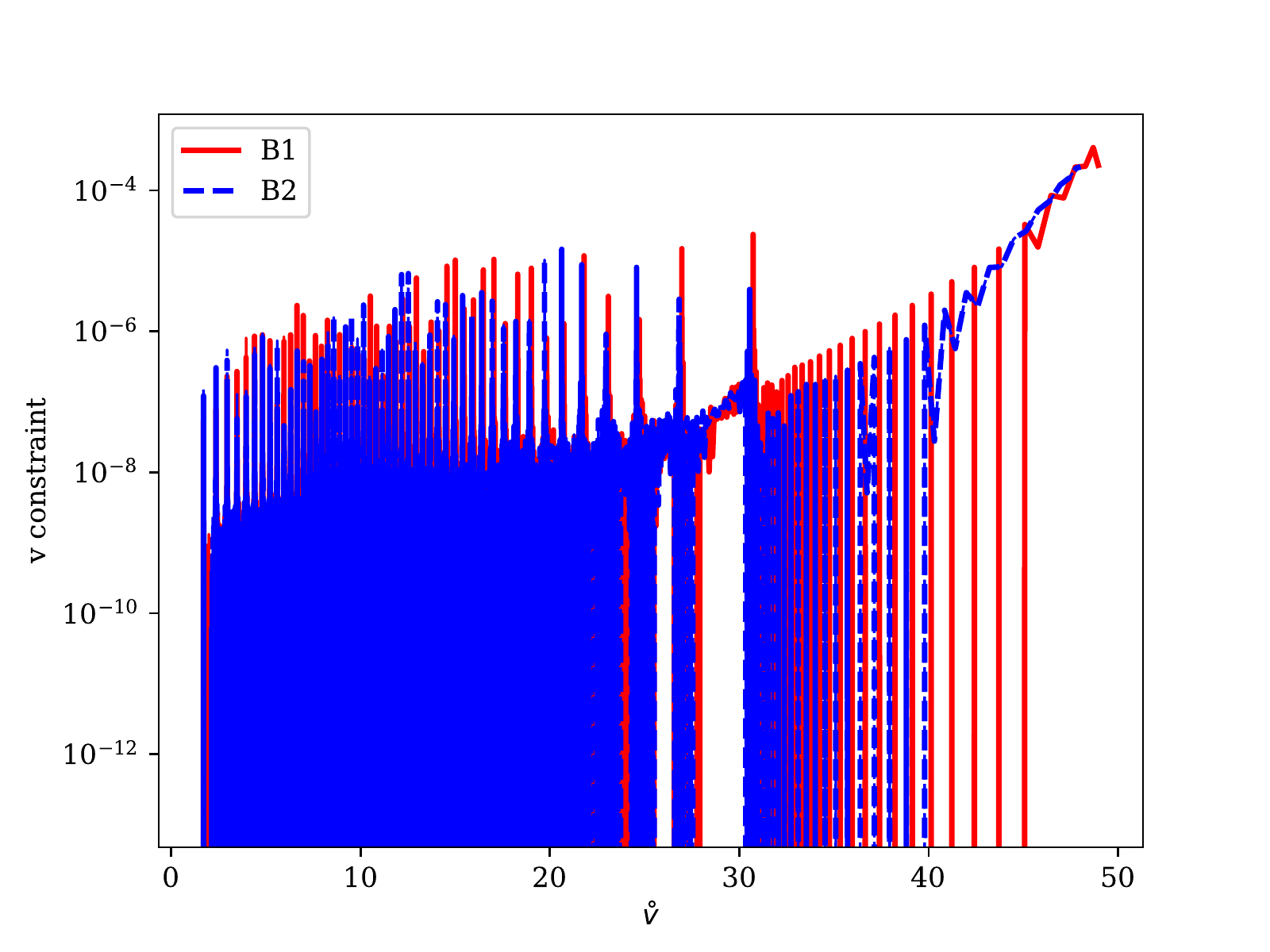}
  \caption{Constraint violations during our evolutions of configurations {\bf
      B}. \label{fig:constraints}}
\end{figure*}

Finally, since we use a free-evolution scheme, we have checked that the
constraint equations~(\ref{eq:constraintu}) and (\ref{eq:constraintv}) remain satisfied throughout our evolution. We show typical plots for the corresponding constraint violation in Fig.~\ref{fig:constraints}.

\section{The Kretschmann scalar}
To avoid round-off errors we use the expression for the Kretschmann
scalar in~\cite{Costa:2014aia} (adapted to include a scalar field mass), instead of computing it directly in terms of the metric:
\begin{widetext}
\beq
K\; &\equiv&\; R_{\alpha \beta \gamma \delta} R^{\alpha \beta \gamma \delta}= \frac{16}{r^6} \left[
      \left(M - \frac{3Q^2}{2r} + \frac{\Lambda}{6}r^3\right)
      + \frac{r}{2} \left(1-\frac{2M}{r} + \frac{Q^2}{r^2} - \frac{\Lambda}{3}r^2\right)  \left(\frac{r \Phi_{,u}}{r_{,u}}\right)
      \left(\frac{r \Phi_{,v}}{r_{,v}} \right)\right]^2 \nonumber\\
    &+& \frac{16}{r^6}\left(M-\frac{Q^2}{2r} + \frac{\Lambda}{6}r^3
    \right)^2+\frac{16}{r^6}\left(M-\frac{Q^2}{r}-\frac{r^3}{3}\left(\Lambda+\mu^2\Phi^2\right)
    \right)^2 
    + \frac{4}{r^4} \left(1-\frac{2M}{r}+\frac{Q^2}{r^2}-\frac{\Lambda}{3}r^2\right)^2  \left(\frac{r \Phi_{,u}}{r_{,u}}\right)^2
      \left(\frac{r \Phi_{,v}}{r_{,v}} \right)^2\,,\nonumber
\eeq
where $M$ is the Misner-Sharp mass.
\end{widetext}
%

\section{The scalar field along the EH}
%
\begin{figure*}[tphb]
  \includegraphics[width=0.3\textwidth]{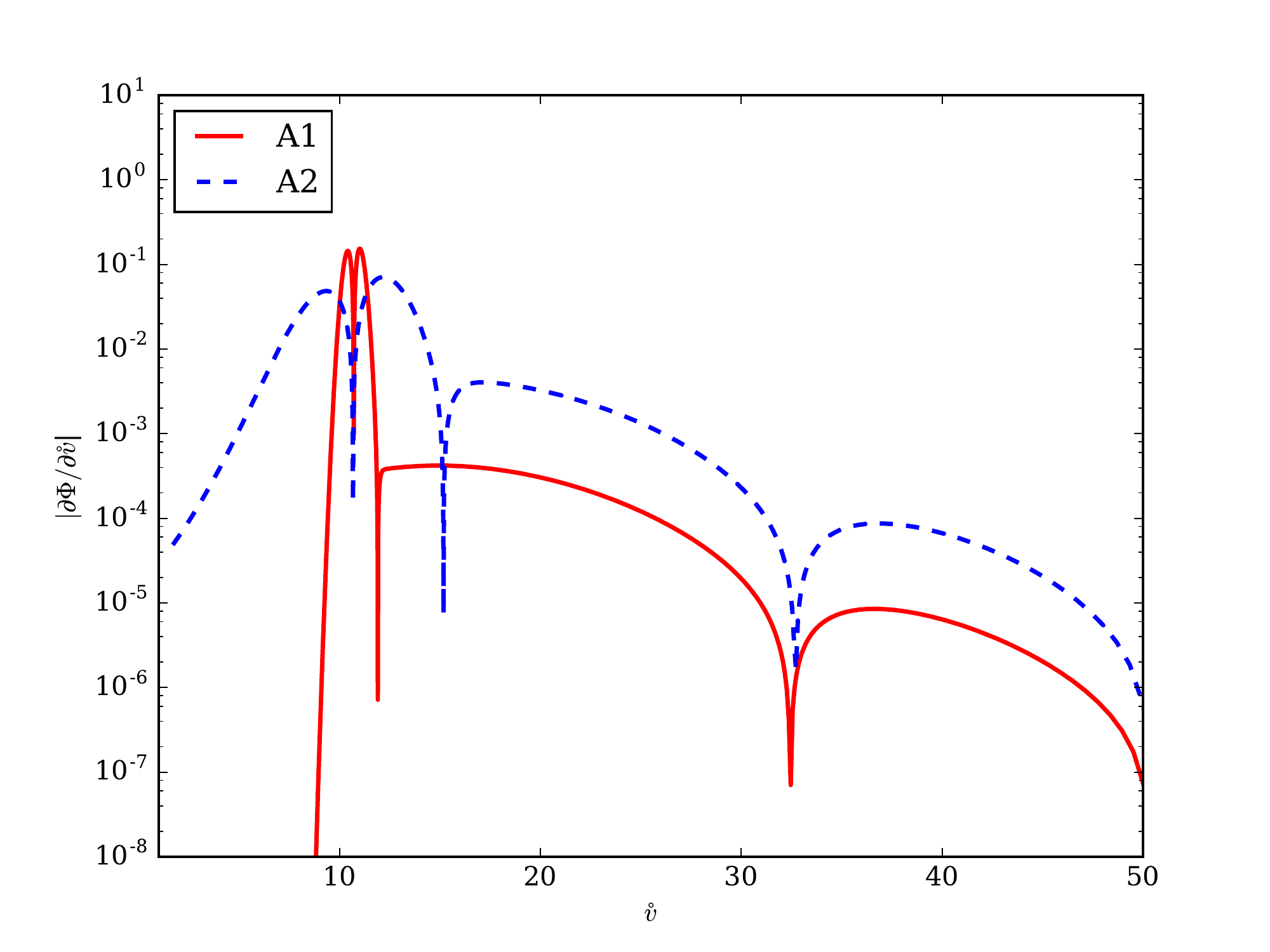}
  \includegraphics[width=0.3\textwidth]{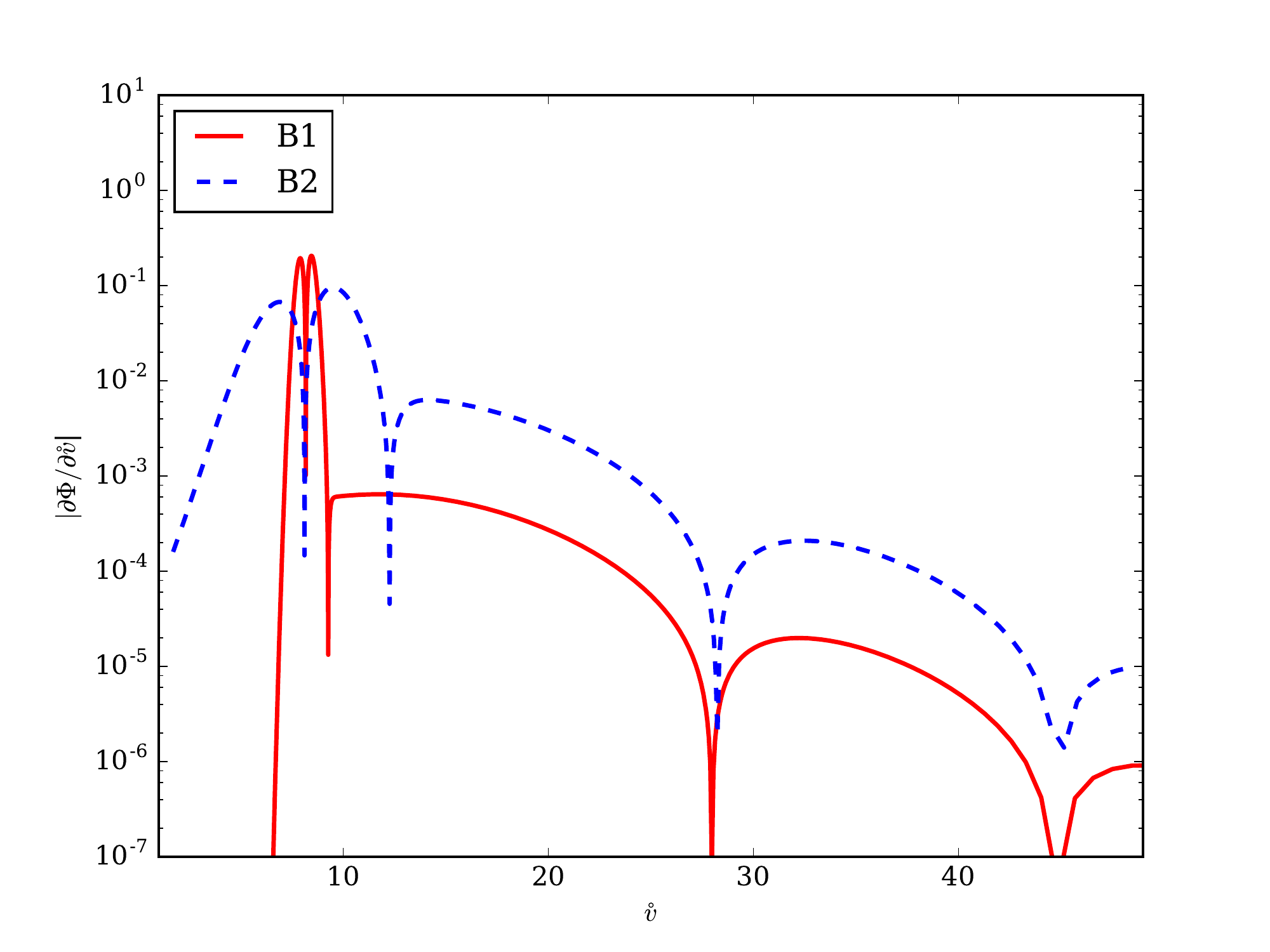}
  \includegraphics[width=0.3\textwidth]{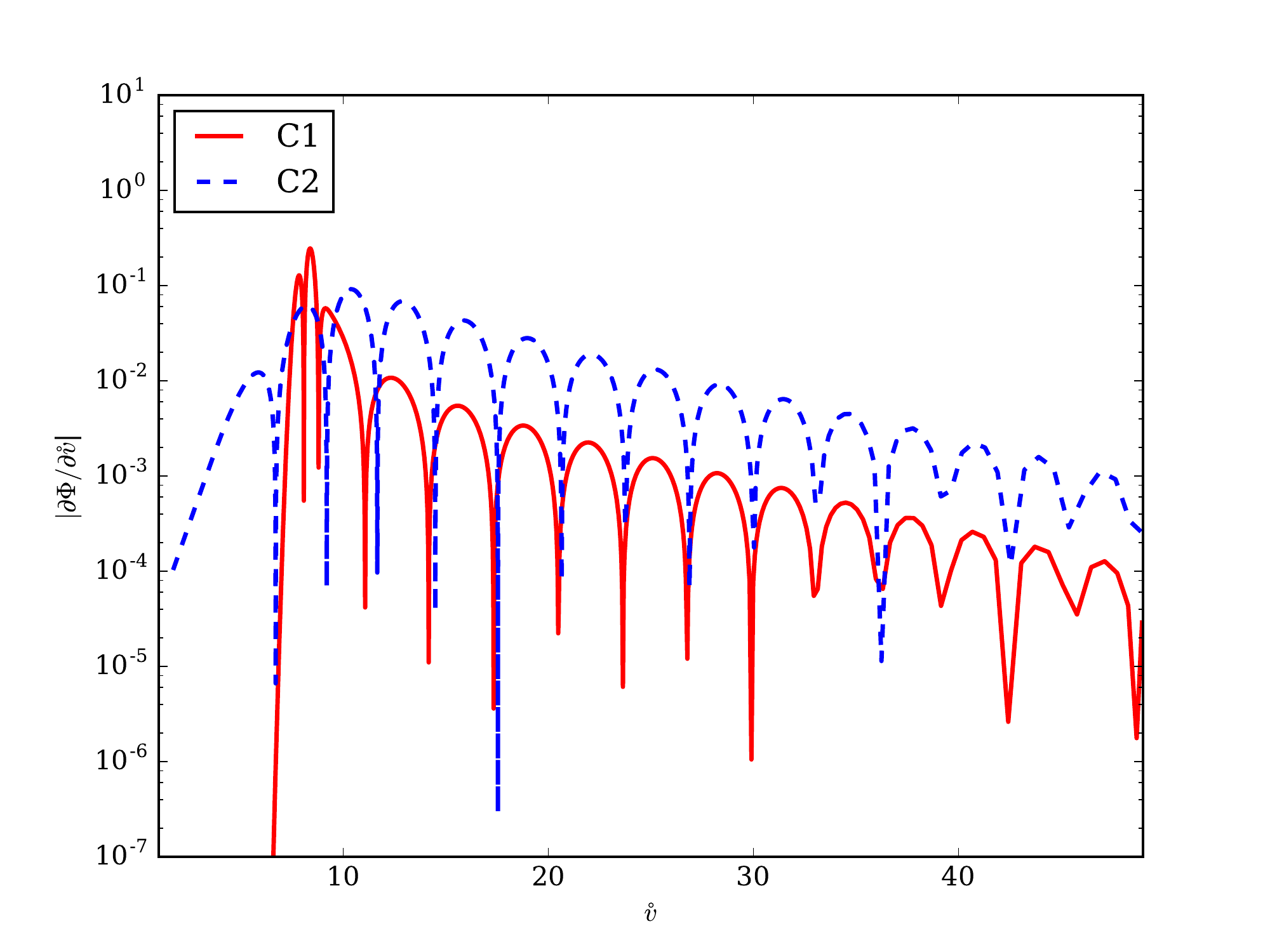}
  \caption{Scalar field derivative $\partial \Phi /\partial \overset{\circ}{v}$ as a function of 
  $\overset{\circ}{v}$ for configurations {\bf A} (left plot), {\bf B} (middle plot), {\bf C} (right plot) with initial
  profiles 1 (red solid lines) and 2 (blue dashed lines). $\partial \Phi /\partial \overset{\circ}{v}$
  evaluated  at $u=u_{\rm EH}$.}
  \label{fig:price}
\end{figure*}
The scalar field along the event horizon is shown in Fig.~\ref{fig:price} (contrast with the asymptotically flat example of Fig.~\ref{fig:latetime}). The late-time behavior is described by ringing exponential falloff of the signal, well described by the lowest quasinormal modes of the spacetime~\cite{Cardoso:2017soq}.

\bibliography{references}

\begin{thebibliography}{32}%
\makeatletter
\providecommand \@ifxundefined [1]{%
 \@ifx{#1\undefined}
}%
\providecommand \@ifnum [1]{%
 \ifnum #1\expandafter \@firstoftwo
 \else \expandafter \@secondoftwo
 \fi
}%
\providecommand \@ifx [1]{%
 \ifx #1\expandafter \@firstoftwo
 \else \expandafter \@secondoftwo
 \fi
}%
\providecommand \natexlab [1]{#1}%
\providecommand \enquote  [1]{``#1''}%
\providecommand \bibnamefont  [1]{#1}%
\providecommand \bibfnamefont [1]{#1}%
\providecommand \citenamefont [1]{#1}%
\providecommand \href@noop [0]{\@secondoftwo}%
\providecommand \href [0]{\begingroup \@sanitize@url \@href}%
\providecommand \@href[1]{\@@startlink{#1}\@@href}%
\providecommand \@@href[1]{\endgroup#1\@@endlink}%
\providecommand \@sanitize@url [0]{\catcode `\\12\catcode `\$12\catcode
  `\&12\catcode `\#12\catcode `\^12\catcode `\_12\catcode `\%12\relax}%
\providecommand \@@startlink[1]{}%
\providecommand \@@endlink[0]{}%
\providecommand \url  [0]{\begingroup\@sanitize@url \@url }%
\providecommand \@url [1]{\endgroup\@href {#1}{\urlprefix }}%
\providecommand \urlprefix  [0]{URL }%
\providecommand \Eprint [0]{\href }%
\providecommand \doibase [0]{http://dx.doi.org/}%
\providecommand \selectlanguage [0]{\@gobble}%
\providecommand \bibinfo  [0]{\@secondoftwo}%
\providecommand \bibfield  [0]{\@secondoftwo}%
\providecommand \translation [1]{[#1]}%
\providecommand \BibitemOpen [0]{}%
\providecommand \bibitemStop [0]{}%
\providecommand \bibitemNoStop [0]{.\EOS\space}%
\providecommand \EOS [0]{\spacefactor3000\relax}%
\providecommand \BibitemShut  [1]{\csname bibitem#1\endcsname}%
\let\auto@bib@innerbib\@empty
\bibitem [{\citenamefont {Costa}\ \emph {et~al.}(2018)\citenamefont {Costa},
  \citenamefont {Gir\~ao}, \citenamefont {Nat\'ario},\ and\ \citenamefont
  {Silva}}]{Costa:2017tjc}%
  \BibitemOpen
  \bibfield  {author} {\bibinfo {author} {\bibfnamefont {J.~L.}\ \bibnamefont
  {Costa}}, \bibinfo {author} {\bibfnamefont {P.~M.}\ \bibnamefont {Gir\~ao}},
  \bibinfo {author} {\bibfnamefont {J.}~\bibnamefont {Nat\'ario}}, \ and\
  \bibinfo {author} {\bibfnamefont {J.~D.}\ \bibnamefont {Silva}},\ }\href
  {\doibase 10.1007/s00220-018-3122-z} {\bibfield  {journal} {\bibinfo
  {journal} {Commun. Math. Phys.}\ }\textbf {\bibinfo {volume} {361}},\
  \bibinfo {pages} {289} (\bibinfo {year} {2018})},\ \Eprint
  {http://arxiv.org/abs/1707.08975} {arXiv:1707.08975 [gr-qc]} \BibitemShut
  {NoStop}%
\bibitem [{\citenamefont {Costa}\ and\ \citenamefont
  {Franzen}(2017)}]{Costa:2016afl}%
  \BibitemOpen
  \bibfield  {author} {\bibinfo {author} {\bibfnamefont {J.~L.}\ \bibnamefont
  {Costa}}\ and\ \bibinfo {author} {\bibfnamefont {A.~T.}\ \bibnamefont
  {Franzen}},\ }\href {\doibase 10.1007/s00023-017-0592-z} {\bibfield
  {journal} {\bibinfo  {journal} {Annales Henri Poincare}\ }\textbf {\bibinfo
  {volume} {18}},\ \bibinfo {pages} {3371} (\bibinfo {year} {2017})},\ \Eprint
  {http://arxiv.org/abs/1607.01018} {arXiv:1607.01018 [gr-qc]} \BibitemShut
  {NoStop}%
\bibitem [{\citenamefont {Hintz}\ and\ \citenamefont
  {Vasy}(2017)}]{Hintz:2015jkj}%
  \BibitemOpen
  \bibfield  {author} {\bibinfo {author} {\bibfnamefont {P.}~\bibnamefont
  {Hintz}}\ and\ \bibinfo {author} {\bibfnamefont {A.}~\bibnamefont {Vasy}},\
  }\href {\doibase 10.1063/1.4996575} {\bibfield  {journal} {\bibinfo
  {journal} {J. Math. Phys.}\ }\textbf {\bibinfo {volume} {58}},\ \bibinfo
  {pages} {081509} (\bibinfo {year} {2017})},\ \Eprint
  {http://arxiv.org/abs/1512.08004} {arXiv:1512.08004 [math.AP]} \BibitemShut
  {NoStop}%
\bibitem [{\citenamefont {Cardoso}\ \emph
  {et~al.}(2018{\natexlab{a}})\citenamefont {Cardoso}, \citenamefont {Costa},
  \citenamefont {Destounis}, \citenamefont {Hintz},\ and\ \citenamefont
  {Jansen}}]{Cardoso:2017soq}%
  \BibitemOpen
  \bibfield  {author} {\bibinfo {author} {\bibfnamefont {V.}~\bibnamefont
  {Cardoso}}, \bibinfo {author} {\bibfnamefont {J.~L.}\ \bibnamefont {Costa}},
  \bibinfo {author} {\bibfnamefont {K.}~\bibnamefont {Destounis}}, \bibinfo
  {author} {\bibfnamefont {P.}~\bibnamefont {Hintz}}, \ and\ \bibinfo {author}
  {\bibfnamefont {A.}~\bibnamefont {Jansen}},\ }\href {\doibase
  10.1103/PhysRevLett.120.031103} {\bibfield  {journal} {\bibinfo  {journal}
  {Phys. Rev. Lett.}\ }\textbf {\bibinfo {volume} {120}},\ \bibinfo {pages}
  {031103} (\bibinfo {year} {2018}{\natexlab{a}})},\ \Eprint
  {http://arxiv.org/abs/1711.10502} {arXiv:1711.10502 [gr-qc]} \BibitemShut
  {NoStop}%
\bibitem [{\citenamefont {Cardoso}\ \emph
  {et~al.}(2018{\natexlab{b}})\citenamefont {Cardoso}, \citenamefont {Costa},
  \citenamefont {Destounis}, \citenamefont {Hintz},\ and\ \citenamefont
  {Jansen}}]{Cardoso:2018nvb}%
  \BibitemOpen
  \bibfield  {author} {\bibinfo {author} {\bibfnamefont {V.}~\bibnamefont
  {Cardoso}}, \bibinfo {author} {\bibfnamefont {J.~L.}\ \bibnamefont {Costa}},
  \bibinfo {author} {\bibfnamefont {K.}~\bibnamefont {Destounis}}, \bibinfo
  {author} {\bibfnamefont {P.}~\bibnamefont {Hintz}}, \ and\ \bibinfo {author}
  {\bibfnamefont {A.}~\bibnamefont {Jansen}},\ }\href@noop {} {\  (\bibinfo
  {year} {2018}{\natexlab{b}})},\ \Eprint {http://arxiv.org/abs/1808.03631}
  {arXiv:1808.03631 [gr-qc]} \BibitemShut {NoStop}%
\bibitem [{\citenamefont {Dias}\ \emph
  {et~al.}(2018{\natexlab{a}})\citenamefont {Dias}, \citenamefont {Reall},\
  and\ \citenamefont {Santos}}]{Dias:2018etb}%
  \BibitemOpen
  \bibfield  {author} {\bibinfo {author} {\bibfnamefont {O.~J.~C.}\
  \bibnamefont {Dias}}, \bibinfo {author} {\bibfnamefont {H.~S.}\ \bibnamefont
  {Reall}}, \ and\ \bibinfo {author} {\bibfnamefont {J.~E.}\ \bibnamefont
  {Santos}},\ }\href@noop {} {\  (\bibinfo {year} {2018}{\natexlab{a}})},\
  \Eprint {http://arxiv.org/abs/1808.02895} {arXiv:1808.02895 [gr-qc]}
  \BibitemShut {NoStop}%
\bibitem [{\citenamefont {Mo}\ \emph {et~al.}(2018)\citenamefont {Mo},
  \citenamefont {Tian}, \citenamefont {Wang}, \citenamefont {Zhang},\ and\
  \citenamefont {Zhong}}]{Mo:2018nnu}%
  \BibitemOpen
  \bibfield  {author} {\bibinfo {author} {\bibfnamefont {Y.}~\bibnamefont
  {Mo}}, \bibinfo {author} {\bibfnamefont {Y.}~\bibnamefont {Tian}}, \bibinfo
  {author} {\bibfnamefont {B.}~\bibnamefont {Wang}}, \bibinfo {author}
  {\bibfnamefont {H.}~\bibnamefont {Zhang}}, \ and\ \bibinfo {author}
  {\bibfnamefont {Z.}~\bibnamefont {Zhong}},\ }\href@noop {} {\  (\bibinfo
  {year} {2018})},\ \Eprint {http://arxiv.org/abs/1808.03635} {arXiv:1808.03635
  [gr-qc]} \BibitemShut {NoStop}%
\bibitem [{\citenamefont {Dias}\ \emph
  {et~al.}(2018{\natexlab{b}})\citenamefont {Dias}, \citenamefont {Reall},\
  and\ \citenamefont {Santos}}]{Dias:2018ufh}%
  \BibitemOpen
  \bibfield  {author} {\bibinfo {author} {\bibfnamefont {O.~J.~C.}\
  \bibnamefont {Dias}}, \bibinfo {author} {\bibfnamefont {H.~S.}\ \bibnamefont
  {Reall}}, \ and\ \bibinfo {author} {\bibfnamefont {J.~E.}\ \bibnamefont
  {Santos}},\ }\href@noop {} {\  (\bibinfo {year} {2018}{\natexlab{b}})},\
  \Eprint {http://arxiv.org/abs/1808.04832} {arXiv:1808.04832 [gr-qc]}
  \BibitemShut {NoStop}%
\bibitem [{\citenamefont {Mellor}\ and\ \citenamefont
  {Moss}(1990)}]{Mellor:1990}%
  \BibitemOpen
  \bibfield  {author} {\bibinfo {author} {\bibfnamefont {F.}~\bibnamefont
  {Mellor}}\ and\ \bibinfo {author} {\bibfnamefont {I.}~\bibnamefont {Moss}},\
  }\href {\doibase 10.1103/PhysRevD.41.403} {\bibfield  {journal} {\bibinfo
  {journal} {Phys. Rev.}\ }\textbf {\bibinfo {volume} {D41}},\ \bibinfo {pages}
  {403} (\bibinfo {year} {1990})}\BibitemShut {NoStop}%
\bibitem [{\citenamefont {Brady}\ \emph {et~al.}(1998)\citenamefont {Brady},
  \citenamefont {Moss},\ and\ \citenamefont {Myers}}]{Brady:1998au}%
  \BibitemOpen
  \bibfield  {author} {\bibinfo {author} {\bibfnamefont {P.~R.}\ \bibnamefont
  {Brady}}, \bibinfo {author} {\bibfnamefont {I.~G.}\ \bibnamefont {Moss}}, \
  and\ \bibinfo {author} {\bibfnamefont {R.~C.}\ \bibnamefont {Myers}},\ }\href
  {\doibase 10.1103/PhysRevLett.80.3432} {\bibfield  {journal} {\bibinfo
  {journal} {Phys. Rev. Lett.}\ }\textbf {\bibinfo {volume} {80}},\ \bibinfo
  {pages} {3432} (\bibinfo {year} {1998})},\ \Eprint
  {http://arxiv.org/abs/gr-qc/9801032} {arXiv:gr-qc/9801032 [gr-qc]}
  \BibitemShut {NoStop}%
\bibitem [{\citenamefont {Dafermos}(2005)}]{Dafermos:2003wr}%
  \BibitemOpen
  \bibfield  {author} {\bibinfo {author} {\bibfnamefont {M.}~\bibnamefont
  {Dafermos}},\ }\href@noop {} {\bibfield  {journal} {\bibinfo  {journal}
  {Commun. Pure Appl. Math.}\ }\textbf {\bibinfo {volume} {58}},\ \bibinfo
  {pages} {0445} (\bibinfo {year} {2005})},\ \Eprint
  {http://arxiv.org/abs/gr-qc/0307013} {arXiv:gr-qc/0307013 [gr-qc]}
  \BibitemShut {NoStop}%
\bibitem [{\citenamefont {Luk}(2018)}]{Luk:2013cqa}%
  \BibitemOpen
  \bibfield  {author} {\bibinfo {author} {\bibfnamefont {J.}~\bibnamefont
  {Luk}},\ }\href {\doibase 10.1090/jams/888} {\bibfield  {journal} {\bibinfo
  {journal} {J. Am. Math. Soc.}\ }\textbf {\bibinfo {volume} {31}},\ \bibinfo
  {pages} {1} (\bibinfo {year} {2018})},\ \Eprint
  {http://arxiv.org/abs/1311.4970} {arXiv:1311.4970 [gr-qc]} \BibitemShut
  {NoStop}%
\bibitem [{\citenamefont {Dafermos}\ and\ \citenamefont
  {Luk}(2017)}]{Dafermos:2017dbw}%
  \BibitemOpen
  \bibfield  {author} {\bibinfo {author} {\bibfnamefont {M.}~\bibnamefont
  {Dafermos}}\ and\ \bibinfo {author} {\bibfnamefont {J.}~\bibnamefont {Luk}},\
  }\href@noop {} {\  (\bibinfo {year} {2017})},\ \Eprint
  {http://arxiv.org/abs/1710.01722} {arXiv:1710.01722 [gr-qc]} \BibitemShut
  {NoStop}%
\bibitem [{\citenamefont {Luk}\ and\ \citenamefont
  {Oh}(2017{\natexlab{a}})}]{Luk:2017jxq}%
  \BibitemOpen
  \bibfield  {author} {\bibinfo {author} {\bibfnamefont {J.}~\bibnamefont
  {Luk}}\ and\ \bibinfo {author} {\bibfnamefont {S.-J.}\ \bibnamefont {Oh}},\
  }\href@noop {} {\  (\bibinfo {year} {2017}{\natexlab{a}})},\ \Eprint
  {http://arxiv.org/abs/1702.05715} {arXiv:1702.05715 [gr-qc]} \BibitemShut
  {NoStop}%
\bibitem [{\citenamefont {Luk}\ and\ \citenamefont
  {Oh}(2017{\natexlab{b}})}]{Luk:2017ofx}%
  \BibitemOpen
  \bibfield  {author} {\bibinfo {author} {\bibfnamefont {J.}~\bibnamefont
  {Luk}}\ and\ \bibinfo {author} {\bibfnamefont {S.-J.}\ \bibnamefont {Oh}},\
  }\href@noop {} {\  (\bibinfo {year} {2017}{\natexlab{b}})},\ \Eprint
  {http://arxiv.org/abs/1702.05716} {arXiv:1702.05716 [gr-qc]} \BibitemShut
  {NoStop}%
\bibitem [{\citenamefont {Van~de Moortel}(2017)}]{VandeMoortel:2017ztd}%
  \BibitemOpen
  \bibfield  {author} {\bibinfo {author} {\bibfnamefont {M.}~\bibnamefont
  {Van~de Moortel}},\ }\href {\doibase 10.1007/s00220-017-3079-3} {\  (\bibinfo
  {year} {2017}),\ 10.1007/s00220-017-3079-3},\ \Eprint
  {http://arxiv.org/abs/1704.05790} {arXiv:1704.05790 [gr-qc]} \BibitemShut
  {NoStop}%
\bibitem [{\citenamefont {Dafermos}(2014)}]{Dafermos:2012np}%
  \BibitemOpen
  \bibfield  {author} {\bibinfo {author} {\bibfnamefont {M.}~\bibnamefont
  {Dafermos}},\ }\href {\doibase 10.1007/s00220-014-2063-4} {\bibfield
  {journal} {\bibinfo  {journal} {Commun. Math. Phys.}\ }\textbf {\bibinfo
  {volume} {332}},\ \bibinfo {pages} {729} (\bibinfo {year} {2014})},\ \Eprint
  {http://arxiv.org/abs/1201.1797} {arXiv:1201.1797 [gr-qc]} \BibitemShut
  {NoStop}%
\bibitem [{\citenamefont {Klainerman}\ \emph {et~al.}(2015)\citenamefont
  {Klainerman}, \citenamefont {Rodnianski},\ and\ \citenamefont
  {Szeftel}}]{L2}%
  \BibitemOpen
  \bibfield  {author} {\bibinfo {author} {\bibfnamefont {S.}~\bibnamefont
  {Klainerman}}, \bibinfo {author} {\bibfnamefont {I.}~\bibnamefont
  {Rodnianski}}, \ and\ \bibinfo {author} {\bibfnamefont {J.}~\bibnamefont
  {Szeftel}},\ }\href {\doibase 10.1007/s00222-014-0567-3} {\bibfield
  {journal} {\bibinfo  {journal} {Invent. Math.}\ }\textbf {\bibinfo {volume}
  {202}},\ \bibinfo {pages} {91} (\bibinfo {year} {2015})},\ \Eprint
  {http://arxiv.org/abs/1204.1767} {arXiv:1204.1767 [math.AP]} \BibitemShut
  {NoStop}%
\bibitem [{\citenamefont {Ori}(1991)}]{Ori:1991zz}%
  \BibitemOpen
  \bibfield  {author} {\bibinfo {author} {\bibfnamefont {A.}~\bibnamefont
  {Ori}},\ }\href {\doibase 10.1103/PhysRevLett.67.789} {\bibfield  {journal}
  {\bibinfo  {journal} {Phys. Rev. Lett.}\ }\textbf {\bibinfo {volume} {67}},\
  \bibinfo {pages} {789} (\bibinfo {year} {1991})}\BibitemShut {NoStop}%
\bibitem [{\citenamefont {Burko}\ and\ \citenamefont
  {Ori}(1997)}]{Burko:1997tb}%
  \BibitemOpen
  \bibfield  {author} {\bibinfo {author} {\bibfnamefont {L.~M.}\ \bibnamefont
  {Burko}}\ and\ \bibinfo {author} {\bibfnamefont {A.}~\bibnamefont {Ori}},\
  }\href {\doibase 10.1103/PhysRevD.56.7820} {\bibfield  {journal} {\bibinfo
  {journal} {Phys. Rev.}\ }\textbf {\bibinfo {volume} {D56}},\ \bibinfo {pages}
  {7820} (\bibinfo {year} {1997})},\ \Eprint
  {http://arxiv.org/abs/gr-qc/9703067} {arXiv:gr-qc/9703067 [gr-qc]}
  \BibitemShut {NoStop}%
\bibitem [{\citenamefont {Hansen}\ \emph {et~al.}(2005)\citenamefont {Hansen},
  \citenamefont {Khokhlov},\ and\ \citenamefont {Novikov}}]{Hansen:2005am}%
  \BibitemOpen
  \bibfield  {author} {\bibinfo {author} {\bibfnamefont {J.}~\bibnamefont
  {Hansen}}, \bibinfo {author} {\bibfnamefont {A.}~\bibnamefont {Khokhlov}}, \
  and\ \bibinfo {author} {\bibfnamefont {I.}~\bibnamefont {Novikov}},\ }\href
  {\doibase 10.1103/PhysRevD.71.064013} {\bibfield  {journal} {\bibinfo
  {journal} {Phys. Rev.}\ }\textbf {\bibinfo {volume} {D71}},\ \bibinfo {pages}
  {064013} (\bibinfo {year} {2005})},\ \Eprint
  {http://arxiv.org/abs/gr-qc/0501015} {arXiv:gr-qc/0501015 [gr-qc]}
  \BibitemShut {NoStop}%
\bibitem [{\citenamefont {Avelino}\ \emph {et~al.}(2011)\citenamefont
  {Avelino}, \citenamefont {Hamilton}, \citenamefont {Herdeiro},\ and\
  \citenamefont {Zilhao}}]{Avelino:2011ee}%
  \BibitemOpen
  \bibfield  {author} {\bibinfo {author} {\bibfnamefont {P.~P.}\ \bibnamefont
  {Avelino}}, \bibinfo {author} {\bibfnamefont {A.~J.~S.}\ \bibnamefont
  {Hamilton}}, \bibinfo {author} {\bibfnamefont {C.~A.~R.}\ \bibnamefont
  {Herdeiro}}, \ and\ \bibinfo {author} {\bibfnamefont {M.}~\bibnamefont
  {Zilhao}},\ }\href {\doibase 10.1103/PhysRevD.84.024019} {\bibfield
  {journal} {\bibinfo  {journal} {Phys. Rev.}\ }\textbf {\bibinfo {volume}
  {D84}},\ \bibinfo {pages} {024019} (\bibinfo {year} {2011})},\ \Eprint
  {http://arxiv.org/abs/1105.4434} {arXiv:1105.4434 [gr-qc]} \BibitemShut
  {NoStop}%
\bibitem [{\citenamefont {Sbierski}(2018)}]{Sbierski:2015nta}%
  \BibitemOpen
  \bibfield  {author} {\bibinfo {author} {\bibfnamefont {J.}~\bibnamefont
  {Sbierski}},\ }\href {\doibase 10.4310/jdg/1518490820} {\bibfield  {journal}
  {\bibinfo  {journal} {J. Diff. Geom.}\ }\textbf {\bibinfo {volume} {108}},\
  \bibinfo {pages} {319} (\bibinfo {year} {2018})},\ \Eprint
  {http://arxiv.org/abs/1507.00601} {arXiv:1507.00601 [gr-qc]} \BibitemShut
  {NoStop}%
\bibitem [{\citenamefont {Brady}\ \emph {et~al.}(1997)\citenamefont {Brady},
  \citenamefont {Chambers}, \citenamefont {Krivan},\ and\ \citenamefont
  {Laguna}}]{Brady:1996za}%
  \BibitemOpen
  \bibfield  {author} {\bibinfo {author} {\bibfnamefont {P.~R.}\ \bibnamefont
  {Brady}}, \bibinfo {author} {\bibfnamefont {C.~M.}\ \bibnamefont {Chambers}},
  \bibinfo {author} {\bibfnamefont {W.}~\bibnamefont {Krivan}}, \ and\ \bibinfo
  {author} {\bibfnamefont {P.}~\bibnamefont {Laguna}},\ }\href {\doibase
  10.1103/PhysRevD.55.7538} {\bibfield  {journal} {\bibinfo  {journal} {Phys.
  Rev.}\ }\textbf {\bibinfo {volume} {D55}},\ \bibinfo {pages} {7538} (\bibinfo
  {year} {1997})},\ \Eprint {http://arxiv.org/abs/gr-qc/9611056}
  {arXiv:gr-qc/9611056 [gr-qc]} \BibitemShut {NoStop}%
\bibitem [{\citenamefont {Poisson}\ and\ \citenamefont
  {Israel}(1989)}]{Poisson:1989zz}%
  \BibitemOpen
  \bibfield  {author} {\bibinfo {author} {\bibfnamefont {E.}~\bibnamefont
  {Poisson}}\ and\ \bibinfo {author} {\bibfnamefont {W.}~\bibnamefont
  {Israel}},\ }\href {\doibase 10.1103/PhysRevLett.63.1663} {\bibfield
  {journal} {\bibinfo  {journal} {Phys. Rev. Lett.}\ }\textbf {\bibinfo
  {volume} {63}},\ \bibinfo {pages} {1663} (\bibinfo {year}
  {1989})}\BibitemShut {NoStop}%
\bibitem [{\citenamefont {Costa}\ \emph {et~al.}(2017)\citenamefont {Costa},
  \citenamefont {Gir\~ao}, \citenamefont {Nat\'ario},\ and\ \citenamefont
  {Silva}}]{Costa:2014aia}%
  \BibitemOpen
  \bibfield  {author} {\bibinfo {author} {\bibfnamefont {J.~L.}\ \bibnamefont
  {Costa}}, \bibinfo {author} {\bibfnamefont {P.~M.}\ \bibnamefont {Gir\~ao}},
  \bibinfo {author} {\bibfnamefont {J.}~\bibnamefont {Nat\'ario}}, \ and\
  \bibinfo {author} {\bibfnamefont {J.~D.}\ \bibnamefont {Silva}},\ }\href
  {\doibase 10.1007/s40818-017-0028-6} {\bibfield  {journal} {\bibinfo
  {journal} {Ann. PDE}\ }\textbf {\bibinfo {volume} {3}} (\bibinfo {year}
  {2017}),\ 10.1007/s40818-017-0028-6},\ \Eprint
  {http://arxiv.org/abs/1406.7261} {arXiv:1406.7261 [gr-qc]} \BibitemShut
  {NoStop}%
\bibitem [{\citenamefont {Dafermos}\ and\ \citenamefont
  {Shlapentokh-Rothman}(2018)}]{Dafermos:2018tha}%
  \BibitemOpen
  \bibfield  {author} {\bibinfo {author} {\bibfnamefont {M.}~\bibnamefont
  {Dafermos}}\ and\ \bibinfo {author} {\bibfnamefont {Y.}~\bibnamefont
  {Shlapentokh-Rothman}},\ }\href@noop {} {\  (\bibinfo {year} {2018})},\
  \Eprint {http://arxiv.org/abs/1805.08764} {arXiv:1805.08764 [gr-qc]}
  \BibitemShut {NoStop}%
\bibitem [{\citenamefont {Dias}\ \emph
  {et~al.}(2018{\natexlab{c}})\citenamefont {Dias}, \citenamefont {Eperon},
  \citenamefont {Reall},\ and\ \citenamefont {Santos}}]{Dias:2018ynt}%
  \BibitemOpen
  \bibfield  {author} {\bibinfo {author} {\bibfnamefont {O.~J.~C.}\
  \bibnamefont {Dias}}, \bibinfo {author} {\bibfnamefont {F.~C.}\ \bibnamefont
  {Eperon}}, \bibinfo {author} {\bibfnamefont {H.~S.}\ \bibnamefont {Reall}}, \
  and\ \bibinfo {author} {\bibfnamefont {J.~E.}\ \bibnamefont {Santos}},\
  }\href {\doibase 10.1103/PhysRevD.97.104060} {\bibfield  {journal} {\bibinfo
  {journal} {Phys. Rev.}\ }\textbf {\bibinfo {volume} {D97}},\ \bibinfo {pages}
  {104060} (\bibinfo {year} {2018}{\natexlab{c}})},\ \Eprint
  {http://arxiv.org/abs/1801.09694} {arXiv:1801.09694 [gr-qc]} \BibitemShut
  {NoStop}%
\bibitem [{\citenamefont {Bezanson}\ \emph {et~al.}(2017)\citenamefont
  {Bezanson}, \citenamefont {Edelman}, \citenamefont {Karpinski},\ and\
  \citenamefont {Shah}}]{bezanson2017julia}%
  \BibitemOpen
  \bibfield  {author} {\bibinfo {author} {\bibfnamefont {J.}~\bibnamefont
  {Bezanson}}, \bibinfo {author} {\bibfnamefont {A.}~\bibnamefont {Edelman}},
  \bibinfo {author} {\bibfnamefont {S.}~\bibnamefont {Karpinski}}, \ and\
  \bibinfo {author} {\bibfnamefont {V.~B.}\ \bibnamefont {Shah}},\ }\href
  {\doibase 10.1137/141000671} {\bibfield  {journal} {\bibinfo  {journal} {SIAM
  review}\ }\textbf {\bibinfo {volume} {59}},\ \bibinfo {pages} {65} (\bibinfo
  {year} {2017})}\BibitemShut {NoStop}%
\bibitem [{\citenamefont {Rackauckas}\ and\ \citenamefont
  {Nie}(2017)}]{rackauckas2017differentialequations}%
  \BibitemOpen
  \bibfield  {author} {\bibinfo {author} {\bibfnamefont {C.}~\bibnamefont
  {Rackauckas}}\ and\ \bibinfo {author} {\bibfnamefont {Q.}~\bibnamefont
  {Nie}},\ }\href {\doibase 10.5334/jors.151} {\bibfield  {journal} {\bibinfo
  {journal} {Journal of Open Research Software}\ }\textbf {\bibinfo {volume}
  {5}} (\bibinfo {year} {2017}),\ 10.5334/jors.151}\BibitemShut {NoStop}%
\bibitem [{\citenamefont {Eilon}\ and\ \citenamefont
  {Ori}(2016)}]{Eilon:2015axa}%
  \BibitemOpen
  \bibfield  {author} {\bibinfo {author} {\bibfnamefont {E.}~\bibnamefont
  {Eilon}}\ and\ \bibinfo {author} {\bibfnamefont {A.}~\bibnamefont {Ori}},\
  }\href {\doibase 10.1103/PhysRevD.93.024016} {\bibfield  {journal} {\bibinfo
  {journal} {Phys. Rev.}\ }\textbf {\bibinfo {volume} {D93}},\ \bibinfo {pages}
  {024016} (\bibinfo {year} {2016})},\ \Eprint
  {http://arxiv.org/abs/1510.05273} {arXiv:1510.05273 [gr-qc]} \BibitemShut
  {NoStop}%
\bibitem [{\citenamefont {Price}(1972)}]{Price:1971fb}%
  \BibitemOpen
  \bibfield  {author} {\bibinfo {author} {\bibfnamefont {R.~H.}\ \bibnamefont
  {Price}},\ }\href {\doibase 10.1103/PhysRevD.5.2419} {\bibfield  {journal}
  {\bibinfo  {journal} {Phys. Rev.}\ }\textbf {\bibinfo {volume} {D5}},\
  \bibinfo {pages} {2419} (\bibinfo {year} {1972})}\BibitemShut {NoStop}%
\end{thebibliography}%

\end{document}